\documentclass[%
 reprint,
 amsmath,amssymb,
 aps,
prd,
showpacs,
floatfix,
]{revtex4-2}

\RequirePackage[colorlinks,citecolor=blue,urlcolor=blue,linkcolor=blue]{hyperref}
\usepackage[caption=false]{subfig}
\usepackage{verbatim}
\usepackage{physics}
\usepackage{float}
\usepackage{graphicx}
\usepackage{dcolumn}
\usepackage{bm}

\begin{document}


\title{Thermal effects in hot and dilute homogeneous asymmetric nuclear matter}

\author{Vishal Parmar}
\author{Manoj K Sharma}%
 \affiliation{School of Physics and Materials Science, Thapar
    Institute of Engineering and Technology, Patiala-147004, India.}

    \email{physics.vishal01@gmail.com}
 \author{S K Patra}
\affiliation{%
Institute of Physics, Bhubaneswar 751005, India \\Homi Bhaba National Institute, Training School Complex, Anushakti Nagar, Mumbai 400 085, India}%

\date{\today}

\begin{abstract}
We present a comprehensive analysis of hot and dilute isospin-asymmetric nuclear matter employing the temperature-dependent effective-relativistic mean-field theory (E-RMF). The E-RMF is  applied to study the effect of $\delta$ and $\omega-\rho$ meson cross-coupling on the thermal properties of asymmetric nuclear matter using two recently developed IOPB-I and G3  parameter sets.  These sets are known to reproduce the nuclear matter properties in agreement with various experimental and observational constraints.   We consider the nuclear matter to be homogeneous and study the equation of state (EoS) for densities, temperature and asymmetry which are relevant for astrophysical simulations such as supernovae explosion.  The effect of temperature is investigated in reference to the  density-dependent free symmetry energy and its higher-order derivatives using the well known parabolic approximation.  The larger value of $\lambda_\omega$ cross-coupling in G3 in addition to the $\delta$ meson coupling in G3 smoothen the free symmetry energy. Thermal effects on various state variables are examined at fixed temperature and isospin asymmetry by separating their T=0 and the finite-T expressions.  The thermal effects are mainly governed by effective mass with larger effective mass estimating larger thermal contribution.    The effect of temperature on isothermal and isentropic incompressibility is discussed which is in harmony with various available microscopic calculations.  The liquid-gas phase transition properties are examined in asymmetric matter with two conserved charges in the context of different slope parameter and comparable symmetry energy in IOPB-I and G3 set. The spinodal instability, binodal curve and critical properties are  found to be influenced by the slope parameter $L_{sym}$. Finally, we consider a more realistic system with the inclusion of electrons and analyse their effect on instability and adiabatic index of isospin asymmetric nuclear matter.

\end{abstract}

                              
\pacs{21.65.+f, 26.60.+c, 26.60.-c, 25.70.Pq }

\maketitle



\section{\label{introduction} Introduction} 
Core-collapse supernova is nature's one of the brightest optical display where million-year life of a giant star ($M > 8M\odot$) is put to an end violently and abruptly within fractions of a second \cite{couch2017mechanism, ccsn}. The exact mechanism of collapse explosion is still not well understood even after several decades of thorough investigations. In recent years, such explosion has been studied using several ab-initio core-collapse simulations where the hydrodynamics equations are solved numerically \cite{Muller2010, ccsnsimulation}.  These simulations estimate that the explosion energy of $\approx$ 10 $^{51}$ erg is attained within the time scale of $\ge$ 1s  \cite{Sawada_2019}.  The temperature of the matter rises to 20 MeV and the density of the bounce can vary up to two times the nuclear saturation density.  The short time scale of collapse does not allow the matter to reach $\beta$ equilibrium and calculations are usually done at a fixed asymmetry $\alpha =\frac{\rho_n-\rho_p}{\rho_n+\rho_p} \approx$ 0.4 \cite{Alam2017,Nishizaki1994}.  

The determination of EoS for isospin-asymmetric nuclear matter (ANM) is relevant in various areas of nuclear physics ranging from finite nuclei to infinite matter.  Not only the understanding of its ground state is important, but  its behaviour at finite temperature is equally significant. The finite temperature behaviour of ANM is relevant in context to astrophysical events such as neutron star mergers, gamma-ray bursts, proto-neutron stars, early universe, etc \cite{Liu_2010}.  Furthermore, the composition of matter inside the neutron star impacts its transportation and cooling process which are governed by the so-called direct URCA process \cite{brown2018}. With the recent detection of gravitational wave (GW170817) which was accompanied by a gamma-ray-burst and electromagnetic afterglow from the merger of a neutron star binary opened a new era of astrophysics \cite{PhysRevLett.119.161101, Abbott_2017}. In view of above, a systematic understanding of asymmetric nuclear matter at finite temperature is highly desirable. 

The nuclear matter which is predominantly governed by a residual short-range strong force and long-range electromagnetic interaction shows various structures which in turn depend upon the parameters such as density, asymmetry, temperature etc.  At low temperature or entropy, the matter is in the non-homogeneous form below the subnuclear density ( $\rho < 0.1$ fm$^{-3}$). The nuclear matter is a mixture of heavy nuclei and lights clusters in the background of nucleon gas \cite{Avancini2008}.  As the density increases, the nuclei become deformed constituting a frustrating system usually known as pasta structures because of the competition between nuclear and electromagnetic forces \cite{grams2017}. Above saturation density, the matter converts to a homogeneous mixture of nucleons which may contain the exotic species such as kaon and pion condensate, hyperons, quark phases etc. The leptons (electrons in low density and electrons with muons at high density) are present in all these structures for charge neutrality. At large-enough temperatures, the inhomogeneous nuclear matter at low density again transforms to the homogeneous phase.  In this work, however, we assume the nuclear matter to be an ideal homogeneous mixture of nucleons. Such investigation of ideal ANM system is significant to understand the underlying qualitative behaviour.  We employ the finite temperature extension of the effective-relativistic mean-field theory (E-RMF) to study the asymmetric nuclear matter at finite temperature.  The recently developed E-RMF parameters namely IOPB-I \cite{iopb} and G3 \cite{g3} are used here to understand various aspects of nuclear matter at finite temperature. These sets have comparable saturation properties at zero temperature but differ in the number of adjustable parameters i.e various self and cross couplings of $\sigma$, $\omega$, $\rho$ and $\delta$ mesons. This provides us with an opportunity to study the effect of these couplings on the finite temperature properties of asymmetric nuclear matter.  

The central motivation of this study is to perform a detailed analysis of the EoS for dilute and hot homogeneous asymmetric nuclear matter within E-RMF formalism. We aim to understand the nuclear matter properties like symmetry energy $F_{sym}$, slope parameter ($L_{sym}$), skewness parameter ($Q_{sym}$), curvature parameter ($K_{sym}$) as a function of temperature.  These are significant properties of asymmetric nuclear matter and  are often used to constraint the EoS around saturation density.  Several finite temperature effects such as thermal effects on various state variable, isothermal and isentropic incompressibility are addressed.   The results presented in this study (below subnuclear density) serve only to differentiate between the  realistic inhomogeneous phase in a supernova and ideal homogeneous phase. This is analogous to the van-der Walls equation of state and  ideal-gas equation in atomic physics. The phase transition property in asymmetric matter in comparison with symmetric matter is discussed in context to incompressibility $K$ and slope parameter $(L_{sym})$. With this study, we  aim to verify the trends available in various studies \cite{Alam2017, SHARMA2020121974, Avancini2006, Avancini2004} where effect of symmetry energy and its derivative is discussed on the instability of ANM. Establishing these trends is of primary importance as they serve as the bridge between various nuclear matter properties which are not measured directly from the experiments. In symmetric nuclear matter the trends are seen among the properties at critical temperature  \cite{vishal2020}. The properties at ground state do not necessarily dictate the critical properties of phase transition. However, for asymmetric matter, the symmetry energy and its slope parameter decides the energetic  and therefore impacts the  instabilities occurring in the system. In a realistic case like supernovae it will affect the transportation and cooling process whereas in neutron star crust, the core-crust boundary becomes the variable of slope parameter $L_{sym}$.

The paper is organized as follows: In Section \ref{formalism}, we give a brief formalism of the E-RMF model at finite temperature. The thermal effects on various thermodynamic functions and stability conditions are mentioned in this section. In Section \ref{results}, we summarise our results where we discuss the model properties in Section \ref{model} and properties at finite temperature in Section \ref{ftp}. We discuss the liquid-gas phase transition in ANM in Section \ref{lgpt}. In Section \ref{electron}, we discuss the influence of electrons on EoS. Finally, we summarise our results in Sect. \ref{conclusion}.

\section{\label{formalism} Formalism}
The Lagrangian and the corresponding energy-density functional of the effective field theory (EFT) motivated E-RMF is documented in literature \cite{Delestal2001, iopb, g3, vishal2020}. The advantage of  E-RMF formalism is that one can ignore the basic difficulties of the formalism, like renormalization and divergence of the system. It takes care of several natural phenomena in ab initio manner which otherwise are absent or are included in ad hoc manner in non-relativistic formalisms. In the effective nuclear field theory, Lagrangian contains the infinite number of terms. The
Lagrangian is expanded in the power of meson fields as a truncation scheme because the
fields have a lower mass compared to nucleon masses. The contribution of each term in
the E-RMF Lagrangian can be calculated by counting the power of expansion. Couplings
present at a particular order cannot be dropped arbitrarily without a proper symmetry
argument. The ambiguity in the expansion is checked by the inclusion of naturalness
constraints and naive dimensional analysis (NDA) . For calibration, the coupling constants
and mass of isoscalar-scalar $\sigma$ meson are fitted to reproduce the experimental values of
saturation density and ground state properties of some known nuclei. This method mocks the result of two- loop contribution in mean-field theory \cite{kumar2020warm}. The basic nucleon-meson E-RMF which involve couplings  of  
$\sigma$, $\omega$, $\rho $ and $\delta $ mesons and the photon with Dirac nucleon upto the fourth order 
is given as \cite{vishal2020}

\begin{widetext}
\begin{eqnarray}
\begin{aligned}
\label{rmftlagrangian}
\mathcal{E}(r)=&\psi^{\dagger}(r)\qty{i\alpha\cdot\grad+\beta[M-\Phi(r)-\tau_3D(r)]+W(r)+\frac{1}{2}\tau_3R(r)+\frac{1+\tau_3}{2} A(r)
-\frac{i\beta \alpha }{2M}\qty(f_\omega \grad W(r)+\frac{1}{2}f_\rho \tau_3 \grad R(r))}\psi(r) \\
&+ \qty(\frac{1}{2}+\frac{k_3\Phi(r)}{3! M}+\frac{k_4}{4!}\frac{\Phi^2(r)}{M^2})\frac{m^2_s}{g^2_s}\Phi(r)^2
-\frac{\zeta_0}{4!}\frac{1}{g^2_\omega}W(r)^4+\frac{1}{2g^2_s}\qty\Big(1+\alpha_1\frac{\Phi(r)}{M})
(\grad \Phi(r))^2\\
&-\frac{1}{2g^2_\omega}\qty\Big(1+\alpha_2\frac{\Phi(r)}{M})(\grad W(r))^2
-\frac{1}{2}\qty\Big(1+\eta_1\frac{\Phi(r)}{M}+\frac{\eta_2}{2}\frac{\Phi^2(r)}{M^2})
\frac{m^2_\omega}{g^2_\omega}W^2(r)-\frac{1}{2e^2}(\grad A^2(r))^2\\
&-\frac{1}{2g^2_\rho}(\grad R(r))^2
-\frac{1}{2}\qty\Big(1+\eta_\rho\frac{\Phi(r)}{M})\frac{m^2_\rho}{g^2_\rho}R^2(r)
-\Lambda_\omega(R^2(r)W^2(r))
+\frac{1}{2g^2_\delta}(\grad D(r))^2+\frac{1}{2}\frac{m^2_\delta}{g^2_\delta}(D(r))^2.
\end{aligned}
\end{eqnarray} 
\end{widetext}
Here $\Phi(r)$, W(r), R(r), D(r) and A(r) are the fields corresponding to $\sigma$, $\omega$, $\rho$ and 
$\delta $ mesons and photon respectively. The $g_s$, $g_{\omega}$, $g_{\rho}$, $g_{\delta}$ and $\frac{e^2}{4\pi }$ 
are the corresponding coupling constants and $m_s$, $m_{\omega}$, $m_{\rho}$ and $m_{\delta}$ are the 
corresponding masses. 
The  zeroth component $T_{00}= H$ and the third component $T_{ii}$ of energy-momentum tensor 

\begin{equation}
\label{set}
T_{\mu\nu}=\partial^\nu\phi(x))\frac{\partial\mathcal{E}}{\partial\partial_\mu \phi(x)}-\eta^{\nu\mu}\mathcal{E},
\end{equation}
yields the energy and pressure density respectively as

\begin{widetext}
\begin{eqnarray}
\begin{aligned}
E=&\sum_{p,n}\frac{\gamma}{(2\pi)^3}\int\dd[3]{k}E^*_{p,n}(k)[n_k(T)+\bar{n}_k(T)]+
\rho W+\qty\Big(\frac{1}{2}+\frac{k_3\Phi}{3! M}+\frac{k_4}{4!}\frac{\Phi^2}{M^2})\frac{m^2_s}{g^2_s}\Phi^2
-\frac{1}{2}\qty\Big(1+\eta_1\frac{\Phi}{M}+\frac{\eta_2}{2}\frac{\Phi^2}{M^2})\frac{m^2_\omega}{g^2_\omega}W^2\\
&-\frac{\zeta_0}{4!}\frac{1}{g^2_\omega}W^4+\frac{1}{2}\rho_3R-\frac{1}{2}\qty\Big(1+
\eta_\rho\frac{\Phi}{M})\frac{m^2_\rho}{g^2_\rho}R^2-\Lambda_\omega(R^2W^2)
+\frac{1}{2}\frac{m^2_\delta}{g^2_\delta}(D)^2.
\end{aligned}
\end{eqnarray}

\begin{eqnarray}
\begin{aligned}
P=&\sum_{p,n}\frac{\gamma}{3(2\pi)^3}\int\dd[3]{k}\frac{k^2}{E^*_{p,n}(k)}[n_k(T)+\bar{n}_k(T)]-
\qty\Big(\frac{1}{2}+\frac{k_3\Phi}{3! M}+\frac{k_4}{4!}\frac{\Phi^2}{M^2})\frac{m^2_s}{g^2_s}\Phi^2
&+\frac{1}{2}\qty\Big(1+\eta_1\frac{\Phi}{M}+\frac{\eta_2}{2}\frac{\Phi^2}{M^2})\frac{m^2_\omega}{g^2_\omega}W^2\\
&+\frac{\zeta_0}{4!}\frac{1}{g^2_\omega}W^4+\frac{1}{2}\qty\Big(1+\eta_\rho\frac{\Phi}{M})\frac{m^2_\rho}{g^2_\rho}R^2
+\Lambda_\omega(R^2W^2)
-\frac{1}{2}\frac{m^2_\delta}{g^2_\delta}(D)^2.
\end{aligned}
\end{eqnarray}
\end{widetext}

In above equations,  $n_k(T)$ and $\bar{n}_k(T)$ are the baryons and antibaryons occupation numbers respectively, which are defined
 by the Fermi distribution function at finite temperature T as\\
\begin{subequations}
 \begin{eqnarray}
n_k(T)&=&\frac{1}{1+\exp\qty(\frac{(E^*(k)-\nu)}{T})},\\
\bar{n}_k(T)&=&\frac{1}{1+\exp\qty(\frac{(E^*(k)+\nu)}{T})},
\end{eqnarray}
\end{subequations}
with E$^*$ as $\sqrt{k^2 + {M^*}^2}$. Here the effective mass is written as \cite{singh2014}
\begin{subequations}
\begin{eqnarray}
\label{effmass}
&M^*_p=M-\Phi(r)-D(r),\\
&M^*_n=M-\Phi(r)+D(r).
\end{eqnarray}
\end{subequations}

The effective chemical potential $\nu$ for proton and neutron  are defined as 
\begin{subequations}
\begin{eqnarray}
\label{effcpot}
\nu_p= \mu - W(r)+\frac{1}{2}R(r),\\
\nu_n= \mu - W(r)-\frac{1}{2}R(r).
\end{eqnarray}
\end{subequations}
The entropy density ($S=s/\rho_b$)  is given as 

 \begin{equation}
\label{entropyeq}
s_i=-2\sum_{i}\int \frac{\dd^3k}{(2\pi)^3} [n_{k} \ln n_{k} + (1-n_{k})\ln(1-n_{k})+(n_{k}\leftrightarrow \bar n_{k})]
\end{equation}

and the free energy is defined as
\begin{equation}
    F=E-TS
\end{equation}
The Free energy density  can be written in the parabolic form of the asymmetry parameter $(\alpha=\frac{\rho_n-\rho_p}{\rho_n+\rho_p})$ as \cite{bednarek2020forms, tsang2009}
\begin{equation}
F(\rho,\alpha,T)=F(\rho,\alpha=0,T)+F_{sym}(\rho,T)\alpha^2
\end{equation}
where $F_{sym}(\rho,T)\alpha^2$ is the free symmetry energy content per
nucleon of the system and $F(\rho,\alpha=0,T)$ is the free energy per
nucleon of symmetric ($\alpha=0$) nuclear matter. The free symmetric energy using the empirical parabolic approximation then can be written as
\begin{equation}
    F_{sym}(\rho,T)=  \frac{F(\rho,T,\alpha=1)}{\rho}-\frac{F(\rho,T,\alpha=0)}{\rho}.
    \label{freesymenergy}
\end{equation}
The free symmetric energy then can be  expanded as a Taylor series around the saturation density $\rho_0$ as
\begin{equation}
\begin{aligned}
\label{fsym}
    F_{sym}(\rho,T)=    F_{sym}(\rho_0,T)+ L_{sym}\mathcal{x} + \frac{K_{sym}}{2!}\mathcal{x}^2+\frac{Q_{sym}}{3!}\mathcal{x}^3+\\O(\mathcal{x}^4),
\end{aligned}
\end{equation}

where $\mathcal{x}=\frac{\rho-\rho_0}{3\rho_0}$ and $L_{sym}$, $K_{sym}$ and $Q_{sym}$ are the slope parameter,
curvature parameter and skewness parameter which are written as
\begin{equation}
    \begin{aligned}
        L_{sym}&=&3\rho\frac{\partial F_{sym}(\rho,T)}{\partial \rho},\\
        K_{sym}&=&9\rho^2\frac{\partial^2 F_{sym}(\rho,T)}{\partial \rho^2},\\
        Q_{sym}&=&27\rho^3\frac{\partial^3 F_{sym}(\rho,T)}{\partial \rho^3}.
    \end{aligned}    
\end{equation}
To infer the effects of finite temperature we focus on the thermal part of the various state variables, that is, the difference between the T=0 and the finite-T expressions for a given thermodynamic function $\mathcal{X}$ \cite{Constantinou2014, Constantinou2015},
\begin{equation}
    \mathcal{X}=\mathcal{X}(\rho,\alpha,T)-\mathcal{X}(\rho,\alpha,0)
\end{equation}
The thermal energy, thermal pressure, thermal free energy density and thermal index  are then written as
\begin{equation}
\label{thermalpart}
    \begin{aligned}
    &E_{th}=E(\alpha,T)-E(\alpha,0)\\
    &P_{th}=P(\alpha,T)-P(\alpha,0)\\
    &F_{th}=F(\alpha,T)-E(\alpha,0)\\
    &\lambda_{th}=1+\frac{E_{th}}{P_{th}}\\
    \end{aligned}
\end{equation}
Thermal contributions to the free symmetry energy is given by
\begin{equation}
F_{sym,th}=F_{sym}(\alpha,T)-E_{sym}(\alpha,0).
\end{equation}
In asymmetric nuclear matter system, there are two conserved charges, baryon number ($\rho_b=\rho_p+\rho_n$) and isospin number ($I_3=I_p+I_n$ ). Therefore, one needs to treat it as a binary system. The system will be stable against separation into two phases if the free energy of a single phase is lower than the free energy in all multi-phase configurations.This is formulated as
\begin{equation}
\mathcal{F}(T,\rho_i)<(1-\lambda)\mathcal{F}(T,\rho_i^{'})+\lambda \mathcal{F}(T,\rho_i^{"}),
\end{equation}
with 
\begin{equation}
\rho_i=(1-\lambda)\rho_i^{'}+ \lambda \rho_i^{"},
\end{equation}
where the two phases are denoted by prime and double prime and $\lambda$ is the volume fraction. In formal terms, stability implies that the free energy density is a convex function of the densities. Convexity implies that the stability against separation into two phases also guarantees stability against separation into an arbitrary number of phases. In other terms, it is necessarily true that the symmetric matrix \cite{Muller1995}
\begin{equation}
\label{symmetricmatrix}
\mathcal{F}_{ij}=\Big(\frac{\partial^2 \mathcal{F}}{\partial\rho_i\partial\rho_j}\Big)_T
\end{equation}
is positive. This results is mechanical and diffusive stability conditions as
\begin{equation}
\label{instabilitycondition}
    \frac{\partial P}{\partial \rho_b}\bigg|_{T,\alpha} > 0 \quad \textrm{and} \quad
    \frac{\partial \mu_p}{\partial \alpha}\bigg|_{T,P} < 0
\end{equation}
If one of the stability conditions is violated, a system with
two phases is energetically favorable. The phase coexistence
is governed by the Gibbs conditions

\begin{equation}
\begin{aligned}
\label{gibbscondition}
\mu_q'T,\rho_b')&=&\mu_q''(T,\rho_b''), (q=n,p)\\
P'(T,\rho_b')&=&P''(T,\rho_b'') 
\end{aligned}
\end{equation}
At the critical points, the pressure, density and temperature are written as $P_c, \rho_c$ and $T_c$. For asymmetric nuclear matter, they are calculated by finding an inflation point at chemical potential isobars as
\begin{equation}
    \frac{\partial \mu_q}{ \partial \alpha}\bigg|_{T=T_c}=\frac{\partial^2 \mu_q}{ \partial \alpha^2}\bigg|_{T=T_c}=0
\end{equation}
Furthermore, one can define isothermal incompressibility of nuclear matter at finite temperature T and asymmetry $\alpha$ as
\begin{equation}
\label{isothermal}
    K^T(\alpha,T)=9\bigg(\rho_b^2\frac{\partial^2 F}{\partial \rho_b^2}\bigg)\bigg|_{\rho_b^T(\alpha,T)} 
\end{equation}
Here,$\rho_b^T$ is the density where free energy has its minimum. The isentropic incompressibility at entropy S and asymmetry $\alpha$ which is an important quantity in supernova collapse is written as \cite{haddad2003}
\begin{equation}
\label{isentropic}
    K^S(\alpha,S)=9\bigg(\rho_b^2\frac{\partial^2 E}{\partial \rho_b^2}\bigg)\bigg|_{\rho_b^T(\alpha,S)}
\end{equation}

\section{\label{results} Results and Discussions }

\subsection{\label{model}  Model Properties}
In this work, we consider two recent E-RMF parameteres, namely, IOPB-I \cite{iopb} and G3 \cite{g3}. The IOPB-I set contains the quartic-order cross-coupling $R(r)^2W(r)^2$ $(\lambda_\omega \ne 0)$ and self-coupling of isoscalar-vector $W(r)^2$ ($\zeta_0$). This set produces the infinite nuclear matter properties at saturation and supersaturation density in consistent with the empirical data. The maximum mass is found to be 2.15 M\textsubscript{\(\odot\)} which satisfy the current GW170817,   observational constraint. The G3 set on the other hand is the most comprehensive parameter that contains all the nucleons and tensor coupling terms in addition to several self and cross coupling components. It is known to estimate neutron-skin thickness over a wide range in harmony with the experimental data. It estimates the  maximum neutron star mass    2.03 M\textsubscript{\(\odot\)} with canonical mass radius of 12.69 km which is a desirable feature in context to observational analysis and finite nuclei experiments. The main feature of the set G3 is that it includes the couplings of nucleons to the $\delta$ and $\rho$ mesons and  cross-couplings of $\sigma$$\omega$ and $\sigma$$\rho$ mesons. The G3 set has positive scalar self couplings $k_3$ and $k_4$ and $\zeta_0$ nearly equal to 1. The G3 set and IOPB-I set also differ in the value of $\lambda_\omega$. IOPB-I has relatively small $\lambda_\omega$ as compared to G3. In Table \ref{bulkproperties}, we present the bulk matter properties of nuclear matter for the G3 and IOPB-I sets and the corresponding empirical values. It is clear that these sets satisfy the well-accepted set of laboratorial, theoretical, and observational constraints.

\begin{figure}
	\centering
		\includegraphics[width=1\linewidth]{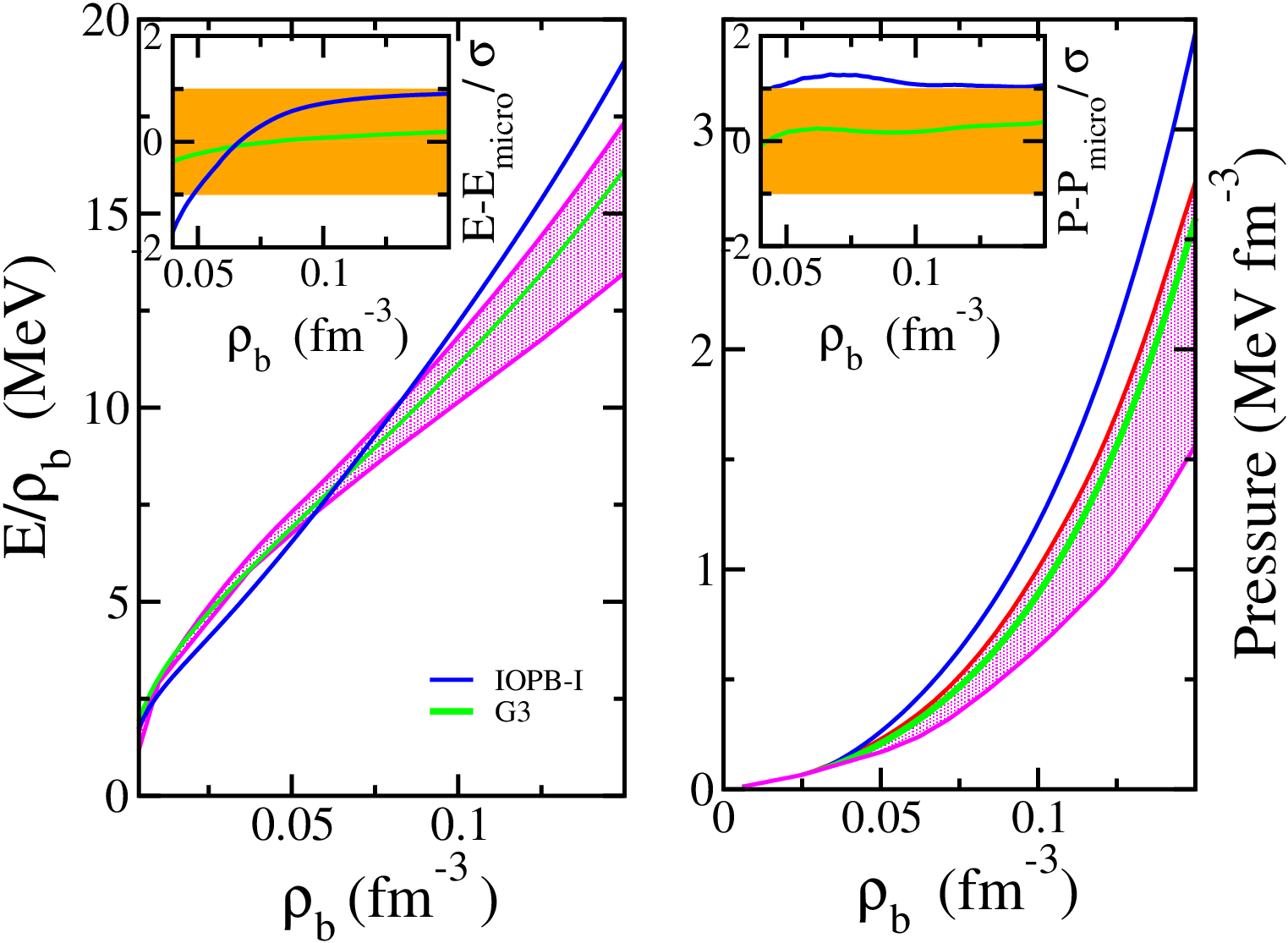}
	\caption{ EoS of nuclear matter below saturation density for pure nuclear matter at T=0 MeV. The shaded magenta region corresponds to the microscopic chiral EFT (NN + 3N) \cite{Hebeler2013} calculations. Inner graph  represents the difference between the neutron matter energy and pressure and the average energy and pressure with 1$\sigma$ calculation uncertainty area.}
	\label{eospnm}
\end{figure}
In Fig \ref{eospnm} we compare the neutron matter binding energy and pressure with the microscopic calculations based on chiral effective field theory (EFT) with realistic two and three-nucleon interactions \cite{Hebeler2013}.  Inner graph  represents the difference between the neutron matter energy and pressure and the average energy and pressure normalised to the  uncertainty of the microscopic calculations($\sigma$=$\delta P$) . The uncertainties are represented by orange band, and they indicate that the points that lie inside this band are within the $1\sigma$ error limits. The G3 set satisfy nicely the microscopic constraints whereas IOPB-I also fall within the $1\sigma$ error limit  below saturation density. These both parameter also satisfy the constraint form collective flow data in heavy-ion collisions and  Kaon experiment along with the GW170817 gravitational wave constraints \cite{iopb}. These features along with the agreement of bulk matter properties with empirical data  motivate us to use them to  study the E-RMF sets with and without the $\delta$ meson. $\delta$ meson couplings are a necessary feature in the dense asymmetric nuclear matter. In this work we intend to investigate the effects of  $\delta$ meson  in the dilute asymmetric matter in the finite temperature limit. We also try to explore role of different self and cross couplings at the finite temperature properties of asymmetric nuclear matter.

\begin{table}[ht]
    \centering
        \caption{Bulk matter properties of nuclear matter for the IOPB-I and G3 parameter and their corresponding empirical values.}
    \begin{tabular*}{\linewidth}{c @{\extracolsep{\fill}}  ccccc}
    \hline
    \hline
    & IOPB-I & G3 & Empirical Value &  \\ \hline
$\rho_0 (fm^{-3})$ & 0.149       &0.148    &    0.148/0.185  \cite{bethe}     &  \\
$E_0$ (MeV)  & -16.10  &-16.02    &       -15.0/-17.0   \cite{bethe}       &  \\
M*/M         &0.593    &    0.699&       0.55/0.6 \cite{marketin2007}    &\\
$J$ (MeV)  &33.30        & 31.84   &        30.0/33.70    \cite{DANIELEWICZ20141}     &  \\
$L$ (MeV)  &63.58        & 49.31   &      35.0/70.0      \cite{DANIELEWICZ20141}     &  \\
$K_{sym}$ (MeV)  & -37.09       &-106.07    &   -174.0/-31.0  \cite{zimmerman2020measuring}            &  \\
$Q _{sym}$ (MeV)  &     862.70   &  915.47  &     -494/-10 \cite{cai2017constraints}            &  \\
$K$ (MeV)  & 222.65        &243.96    &   220/260     \cite{GARG201855}         &  \\
    \hline     
    \hline     
    \end{tabular*}
\label{bulkproperties}
\end{table}

\subsection{\label{ftp} Finite temperature properties}

\begin{figure}[h]
	\centering
		\includegraphics[width=1\linewidth]{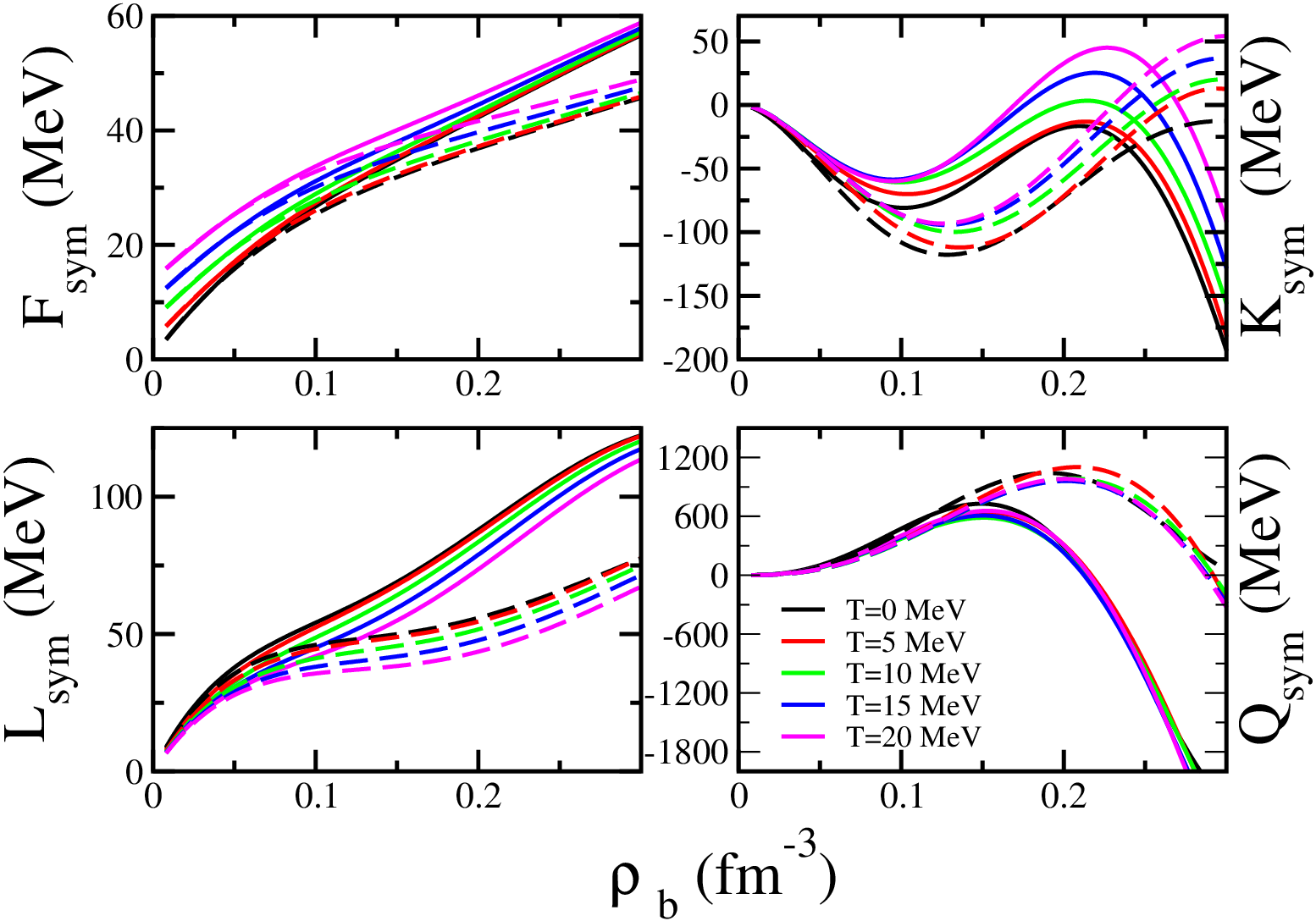}
	\caption{Free symmetry energy $F_{sym.}$, slope parameter L, curvature $K_{sym}$, and isovector skewness parameter $Q_{sym}$  as a function of density at various temperature for IOPB-I (solid lines) and G3 (dashed lines) sets. }
	\label{symenergy}
\end{figure}

\begin{figure*}
  
  \centering
\subfloat[]{%
  \includegraphics[height=8cm,width=.49\linewidth]{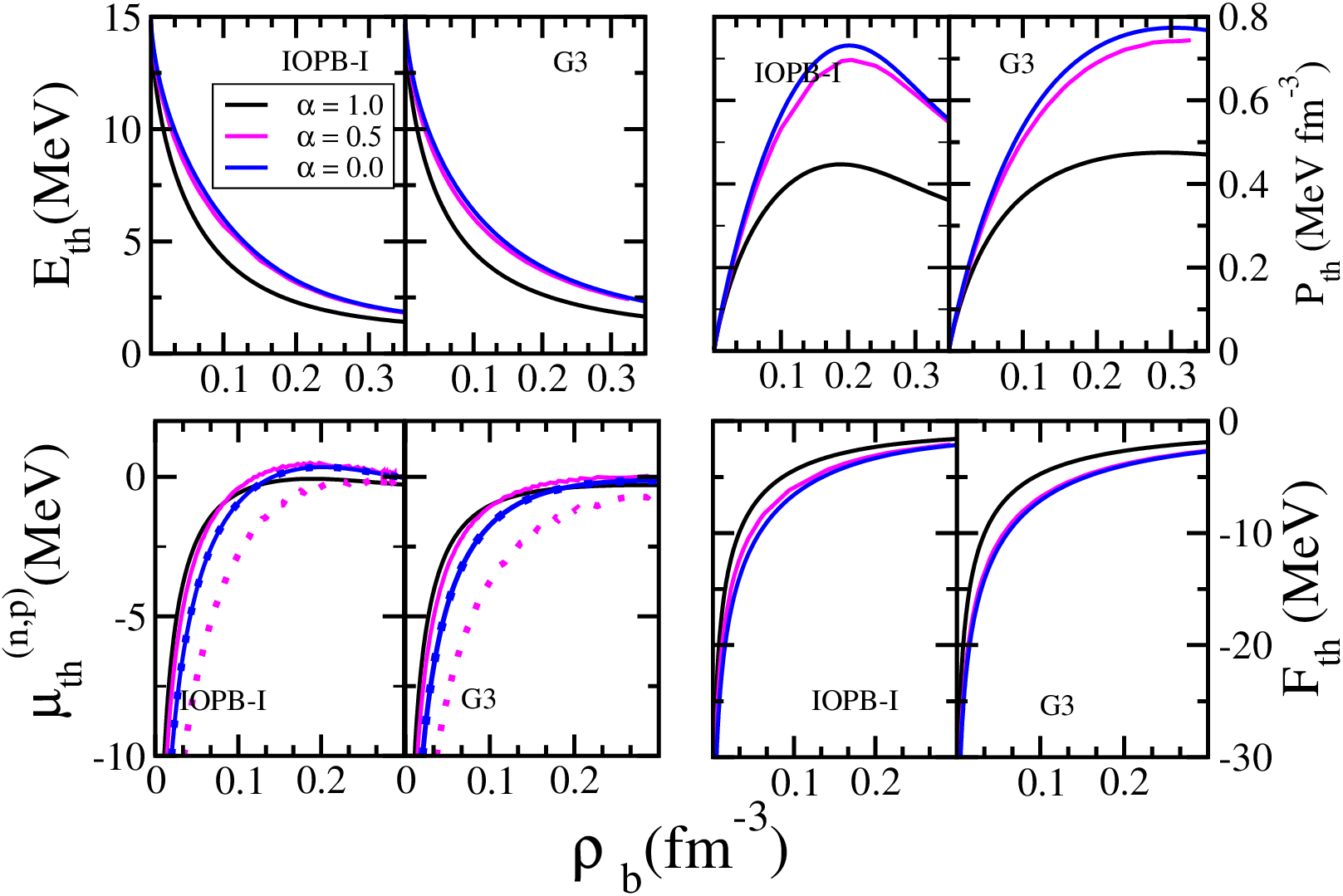}%
}\hfill
\subfloat[]{%
  \includegraphics[height=8cm,width=.49\linewidth]{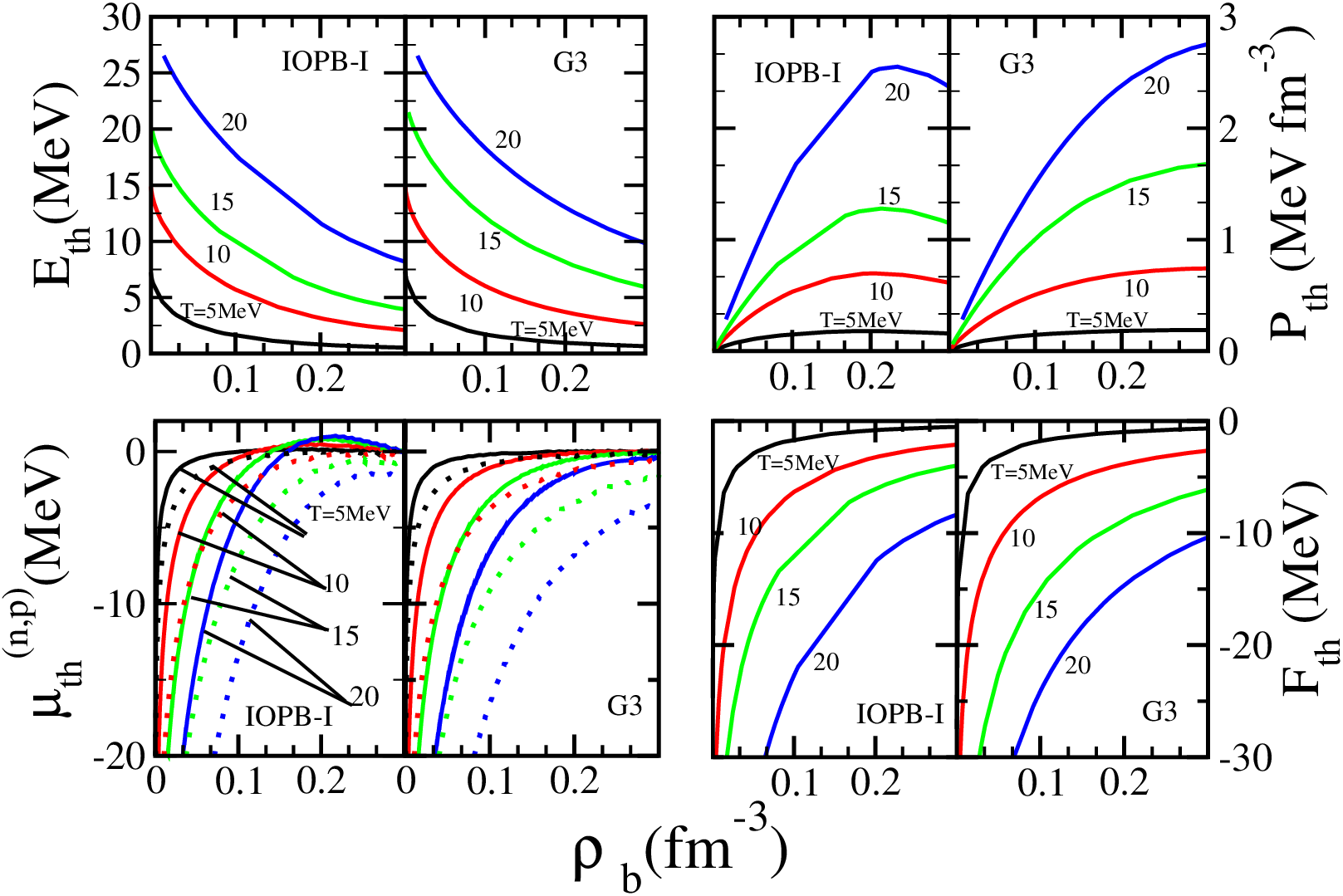}%
}
\caption{(a) Thermal energy, pressure, chemical potential  and free energy density  for fixed T=10 MeV for various $\alpha= 0, 0.5, 1$ (b) same as in (a) but for various T  = 5, 10, 15, 20 MeV at fixed $\alpha=0.5$. The solid lines are for neutron chemical potential and the same colour dotted line represent the proton chemical potential.}
\label{thermaleffect}  
\end{figure*}

The nuclear symmetry energy (NSE) is one of a crucial property of asymmetric nuclear matter governing several areas of nuclear matter calculations like reaction dynamics, phase stability, cooling in the neutron star etc. Temperature dependence of NSE on the other hand provides the much needed knowledge on the dynamical evolution of neutron star and isoscaling analyses of heavy-ion induced  reactions. NSE is not a directly measurable quantity in experiments and is extracted from the observables related
to it. Despite numerous theoretical and experimental efforts, it is still not a very precise parameter even for cold nuclear matter. In Fig. \ref{symenergy}, we show  the variation of Free nuclear symmetry energy (FNSE)  (more relevant quantity in finite temperature case, see Eq. \ref{freesymenergy} ) the slope parameter $L_{sym}$, the isovector incompressibility $K_{sym}$, and the isovector skewness $Q_{sym}$ with the density up to 2 times the saturation density.  The Free NSE is scaled towards higher magnitude due to decrease in entropy preserving its characteristic shape at a higher temperature for both IOPB-I and G3 sets.  At higher density, the temperature range considered here does not affect FSNE much. The slope parameter which has a direct correlation with neutron skin thickness, electric dipole polarizabilities etc also follows the similar trend for both the sets. NSE estimated from these sets are also consistent with the HIC Sn + Sn and IAS data \cite{tsang2009}. The low values of  $F_{sym}$ and $L_{sym}$ are the result of cross-couplings of $\rho$ meson with $\omega$ meson in IOPB-I set and coupling of $\sigma$ meson with $\omega$ meson in G3 set which predicts even lower $F_{sym}$ due to the presence of isovector scalar $\delta$ meson. The $\delta $ meson has the positive effect on binding energy and  helps to estimate the $F_{sym}$, $L_{sym}$ and $ K_{sym}$ within the permissible limit  \cite{singh2014}. The sinusoidal  variation of $K_{sym}$  with density is also shown.   $K_{sym}$  is constrained recently by combining the data from PSR J0030+0451 and GW170817 estimating the $K_{sym}$ = $102^{+71}_{-72}$ MeV within $1\sigma$ error \cite{zimmerman2020measuring} . The IOPB-I and G3 both fall within this constraint. The Variation of $Q_{sym}$ with density is almost independent of temperature for the IOPB-I set and a small variation is observed for G3 set. The $Q_{sym}$ is the least constrained property in any experiment and several models predict it with a large variation \cite{kumar2020warm}.

To study the finite temperature effect, we isolate the thermal part of a given function according to equations \ref{thermalpart}. The subtraction scheme applies to only those variable which depend on the kinetic energy density \cite{Constantinou2015}. Fig \ref{thermaleffect} shows the thermal effect on various state variables at a fixed temperature and $\alpha$ for IOPB-I and G3 parameter set. The common observation is that i) at the fixed temperature, the thermal energy decreases with density. The difference due to asymmetry disappears at high densities and thermal effects become weak and thermal energy gets density independent with asymptotically tending to zero, ii) At very low density, the thermal energy and pressure have linear T dependence as for a free Boltzmann gas (non-degenerate limit). This linearity is changed when matter becomes increasingly degenerate, iii) Temperature effects are more prominent in thermal pressure as compared to thermal energy and iv) The thermal chemical potential becomes saturated after saturation density.  In reference \cite{Constantinou2014, Constantinou2015},  the thermal effects are found to be dominated by the behaviour of effective mass. These calculations were done for Skyrme and APR forces where the effective mass is Landau effective mass whereas in E-RMF the effective mass is Dirac effective mass. For the consistency, one can establish the relationship between both Dirac and Landau mass in relativistic model with constant couplings as $M^*_{Landau,(n,p)}= \sqrt{k^2+M^*_{Dirac, (n,p)}}$ \cite{chen2007}. The qualitative behaviour of both the mass for a given E-RMF set remain same with $M_{Landau}>M_{Dirac}$. In view of this, we can then compare the Dirac mass of E-RMF set with the Landau mass of non-relativistic models. The effective mass mentioned hereafter will therefore be Dirac effective mass only.  

\begin{figure}[h]
	\centering
\includegraphics[width=01\linewidth]{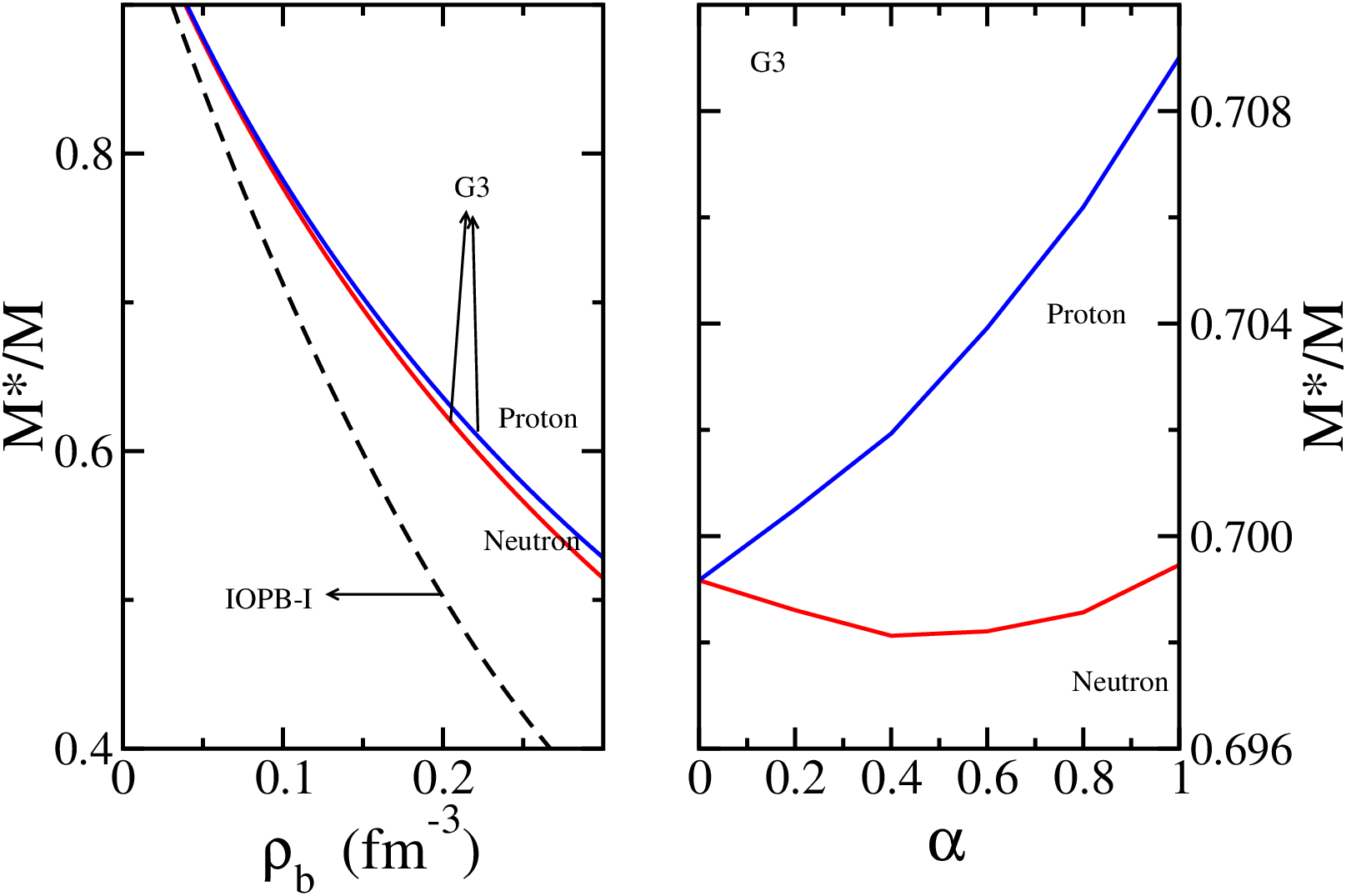}
	\caption{ The effective mass for IOPB-I and G3 set. The left panel shows the mass splitting for PNM in  G3 set (solid lines) as compared to the IOPB-I set (dashed line). In the right panel, the effective mass at saturation density is shown for the G3 set.}
	\label{effectivemass}
	
\end{figure}

In Fig \ref{effectivemass} the density dependence of Dirac effective mass for PNM is shown in the left panel. The effective mass for G3 decreases at a relatively slower pace   as compared to IOPB-I set. Due to the presence of $\delta$ meson, neutron and proton mass gets split which is not the case for IOPB-I set due to the absence of $\delta$ meson. In right panel, the effective mass at saturation density is plotted for different values of $\alpha$ for G3 parameter set. This $\delta$ meson mechanism on effective mass is an in important phenomenon in studying the drip line nuclei which are of astrophysical interests \cite{Horowitz2001} and is analysed in experiments such as PREX \cite{prex}. The effective mass is the input for the computation of energy, pressure and chemical potential which is determined self consistently.  The behaviour of effective mass  therefore clearly dictates the thermal pressure and thermal energy. The G3 set with larger effective mass estimates greater thermal contribution on state variables as compared to the IOPB-I set with smaller effective mass. This is in consistent with the Fermi-liquid theory  and non-relativistic calculations \cite{Constantinou2014}.  For IOPB-I set thermal pressure increases and then decreases beyond the saturation density but for G3 it gets saturated at higher density.  Furthermore, the quantitative difference in thermal energy and pressure between IOPB-I and G3 set is due to the difference in the self-coupling of isoscalar-scalar $\sigma$ meson which is responsible for 3N interaction that plays an important role in determining the thermal pressure and energy. These behaviour are analogous to chiral 2N and 3N interaction, although with larger thermal contribution as compared to the many-body self-consistent Green’s function method \cite{Carbone2019}. The decrease in thermal pressure  after reaching to its maximum is the combining effect of incompressibility of EoS at zero temperature and how rapid is the finite temperature pressure. 

The chemical potential at fixed T=10 MeV has interesting behaviour. $\mu_n$ at $\alpha=1$ is crossed over by  $\mu_n$ at $\alpha<1$ with increasing  density while that is not the case for $\mu_p$. Moreover, the crossings of $\mu_n$ occur at a higher density at larger temperature. Chemical potential becomes saturated at higher density because of the increasing degeneracy at higher density. The different nature of chemical potential is again the consequence of effective mass along with the self and cross-coupling of $\sigma$ meson. The thermal free energy tends to zero with increasing density like the chemical potential.
Comparing IOPB-I and G3 sets for thermal properties, it is seen that the parameter set G3 has additional $\delta$ meson coupling whose contribution increases with density. This contribution directly impacts the effective mass (see Eq. \ref{effmass}) which in turn decides the behaviour of various variable studied above. The $\delta$ meson along with the $\sigma$ meson therefore has the direct contribution in thermal properties of EoS.  

\begin{figure}
	\centering
		\includegraphics[width=1\linewidth]{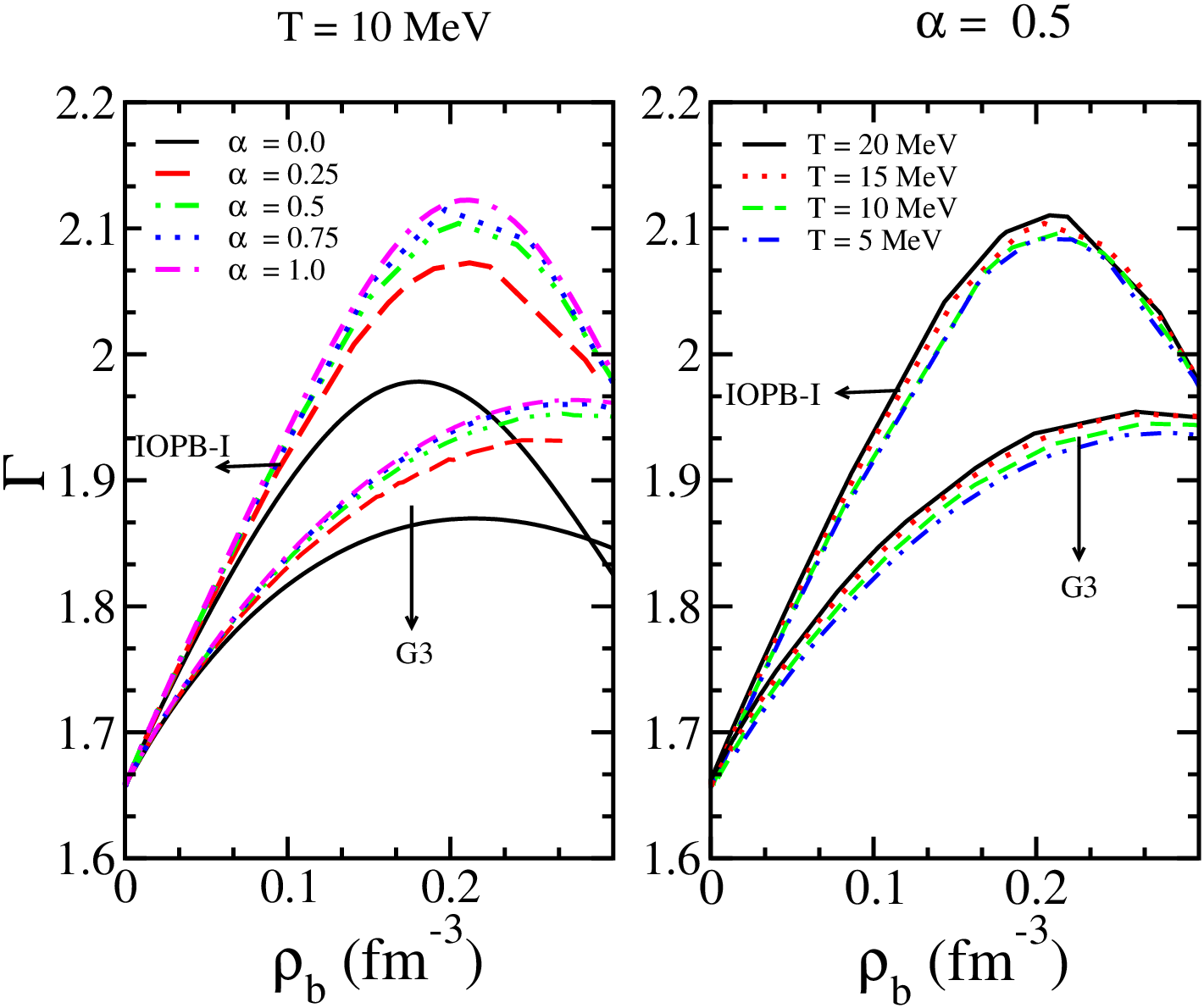}
	\caption{Thermal index for IOPB-I (solid lines) and G3 set (dash line) for fixed temperature in the left panel and fixed $\alpha$ =0.5 in the right panel. }
	\label{thermalindex}
\end{figure}

Fig \ref{thermalindex} shows the variation of the thermal index ($\Gamma$) with density for  IOPB-I and G3 set at fixed temperature and asymmetry. By comparing  the case of fixed temperature and $\alpha$ with those of thermal energy and pressure (Fig. \ref{thermaleffect}) it is certain that $\Gamma$ depends mainly on i) the stiffness of pressure ii) behaviour of effective mass with respect to density and iii) $\alpha$.  For $\rho \rightarrow 0$, $\Gamma$ approaches the  nonrelativistic ideal gas index $\frac{5}{3}$. $\Gamma$ is very sensitive to the asymmetry at a fixed temperature which is opposite to the non-relativistic calculations where the peak of $\Gamma$ is insensitive to asymmetry \cite{Constantinou2015}. Furthermore, it is  immune to temperature change for fixed $\alpha$. For IOPB-I set the maximum $\Gamma$ is 2.1 for PNM while 1.97 for SNM. The G3 set reports these values to be 1.96 and 1.87 respectively. The G3 set with larger effective mass estimate the lower pressure and therefore larger thermal index as compared to the IOPB-I set with lower effective mass. These results of $\Gamma$ from the newly developed E-RMF sets are in consistent with the dynamics of the neutron star merger where $\Gamma$ is taken 1.5 and 2 indicating that these two sets can be used for the calculations of such event \cite{yasin2020}.  However, it is to be noted that, in the astrophysical simulation like binary star and proto-neutron star, $\Gamma$ is taken as a constant while here it varies with the density.  The behaviour of $\Gamma$ is in agreement as with EFT theory \cite{Carbone2019}. 

\begin{figure*}
  \centering
\subfloat[]{%
  \includegraphics[height=8cm,width=.49\linewidth]{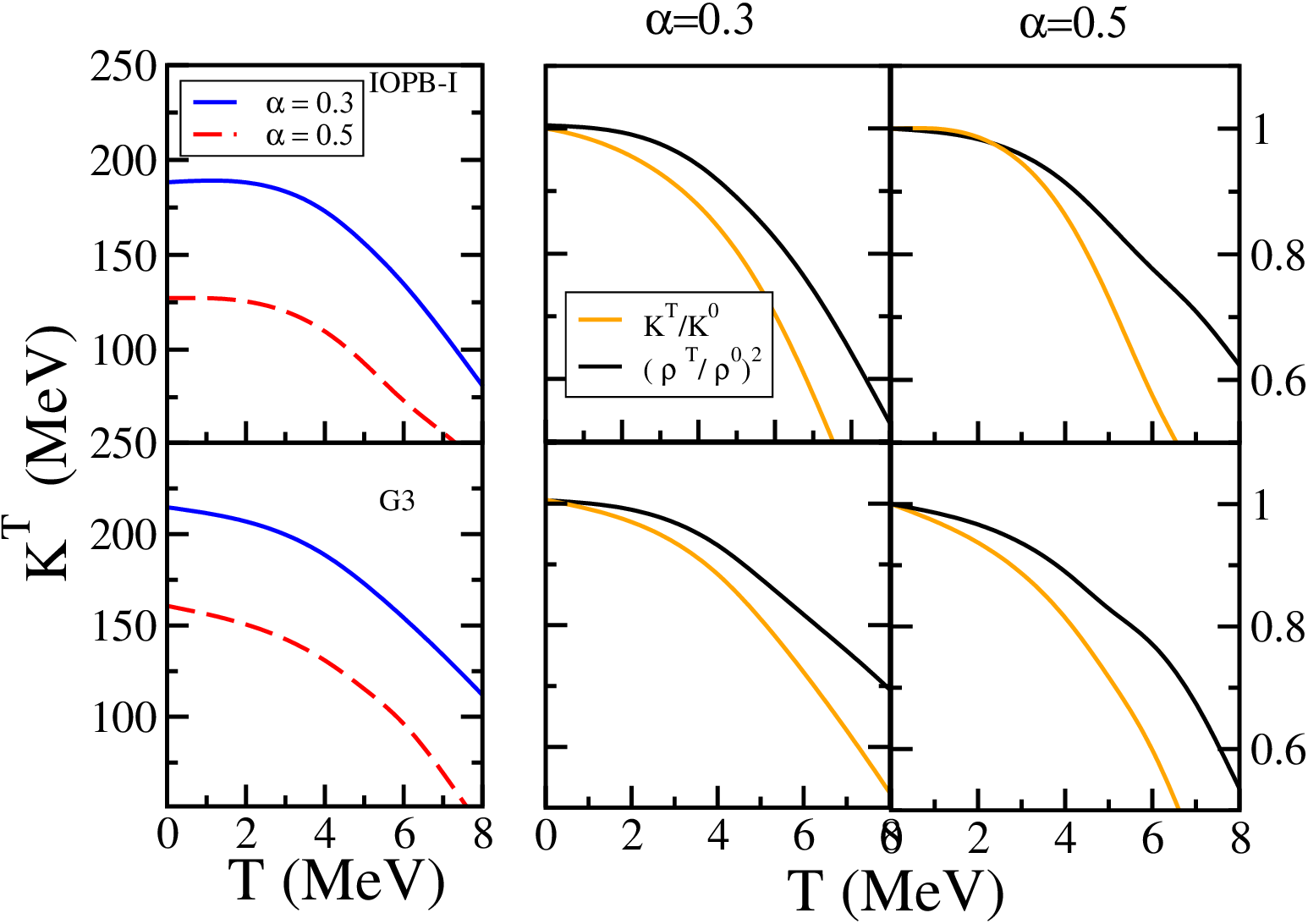}%
}\hfill
\subfloat[]{%
  \includegraphics[height=8cm,width=.49\linewidth]{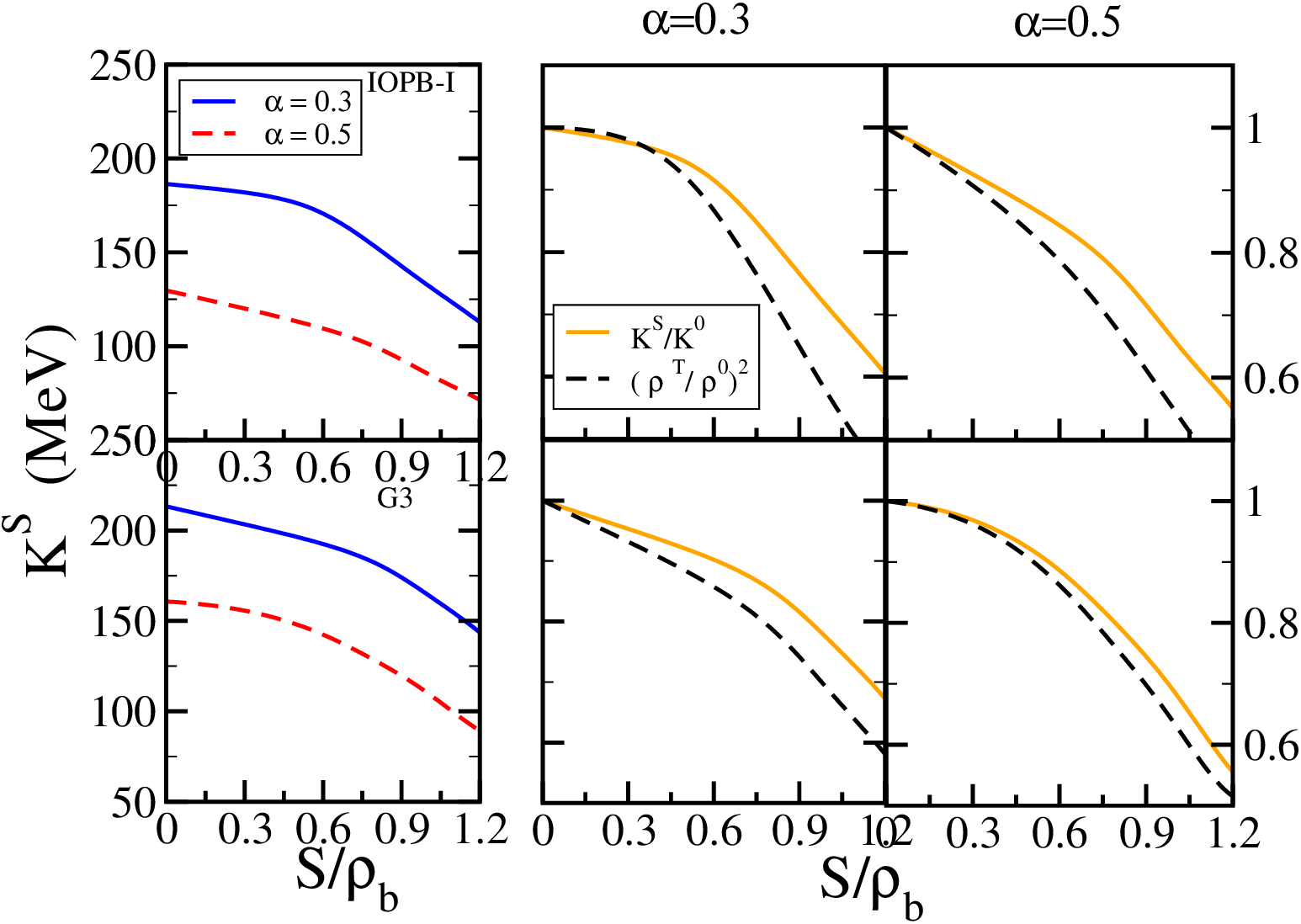}%
} 
\caption{(a) Isothermal and (b) Isentropic incompressibility at $\alpha$ = 0.3 and 0.5 for IOPB-I and G3 set at saturation density.}
  \label{incompressibility}
\end{figure*}

Nuclear matter incompressibility along the isothermal ($K^T$) and isentropic ($K^S$) paths is shown in Fig \ref{incompressibility} for IOPB-I and G3 set for two values of $\alpha$ i.e. 0.3 and 0.5. These values are taken due to their relevance in a core-collapse supernova. The incompressibility at the saturation in the cold nuclear matter is governed` by self-coupling of $\sigma$ meson.  The incompressibility of both IOPB-I and G3 set fall within the accepted empirical value as prescribed by the giant monopole resonance i.e. 240 $\pm$ 20 MeV. At finite temperature, we define incompressibility within two channels. One being the isothermal and other isentropic incompressibility defined according to eq. \ref{isothermal} and \ref{isentropic} respectively. The isentropic incompressibility is more relevant quantity in context to supernova explosion as the time scale of collapse is less than 1 second and the process is adiabatic instead of isothermal. It prompts us to use energy instead of free energy ( see eq  \ref{isentropic}). The incompressibility (both isothermal and isentropic) decreases  quadratically with temperature with G3 having higher magnitude at each temperature and entropy. It also decreases with increasing asymmetry.  We  show the temperature dependence of $K^{T,S}/K^0$ and $(\rho^{T,S}/\rho^0)^2$ in context to their relation with respective incompressibility ( see Eq. \ref{isothermal} and \ref{isentropic}.) Their behaviour remains almost similar irrespective of change in asymmetry. These results satisfy the calculations carried out using microscopic approaches \cite{modarres1998, BOMBACI19939} thereby suggesting that these newly developed parameters not only  describe finite nuclei and cold nuclear matter but can also be used in studying the phenomenon at finite temperature such as proto-neutron star and supernova explosion.  

\subsection{ \label{lgpt} Liquid-gas phase transition}
The asymmetric nuclear matter is a two-component system with two conserved charges Q (B, $I_3$). In a two-component system, although the total charge remains conserved, their ratio can be different in different phases. The constraint on T, Q, and $\rho$ which determine the energetic of the system, forces vapour pressure and chemical potential to change during the phase transition. Apart from mechanical instability, the diffusive instability (fluctuations on the charge concentration) appears and is more relevant to describe the asymmetric matter. The phase transition in the asymmetric matter is therefore described by the following three regions:
\begin{enumerate}
    \item Isothermal Spinodal (ITS): describe the mechanical instability given by $\frac{\partial P}{\partial \rho_b}$. It defines the critical temperature in the symmetric matter.
    \item Diffusive Spinodal (DS):  describe the chemical instability. It essentially means that energy is required to add extra protons in the system at a fixed temperature and pressure. The The critical isobar $P_c$ is estimated by finding a inflation point  $\frac{\partial \mu_p}{\partial \alpha}\Big|_{P_c}$= 0. The corresponding T=$T_c$ and $\rho = \rho_c$ are called the critical temperature and density respectively. 
    \item Coexistence Curve (CE): Set of points where eq. \ref{instabilitycondition} along with the Gibbs conditions are satisfied. This curve may contain the critical points. Unlike symmetric matter case, here CE or binodal is 2 dimensional.
\end{enumerate}
\begin{figure}
	\centering
		\includegraphics[width=1\linewidth]{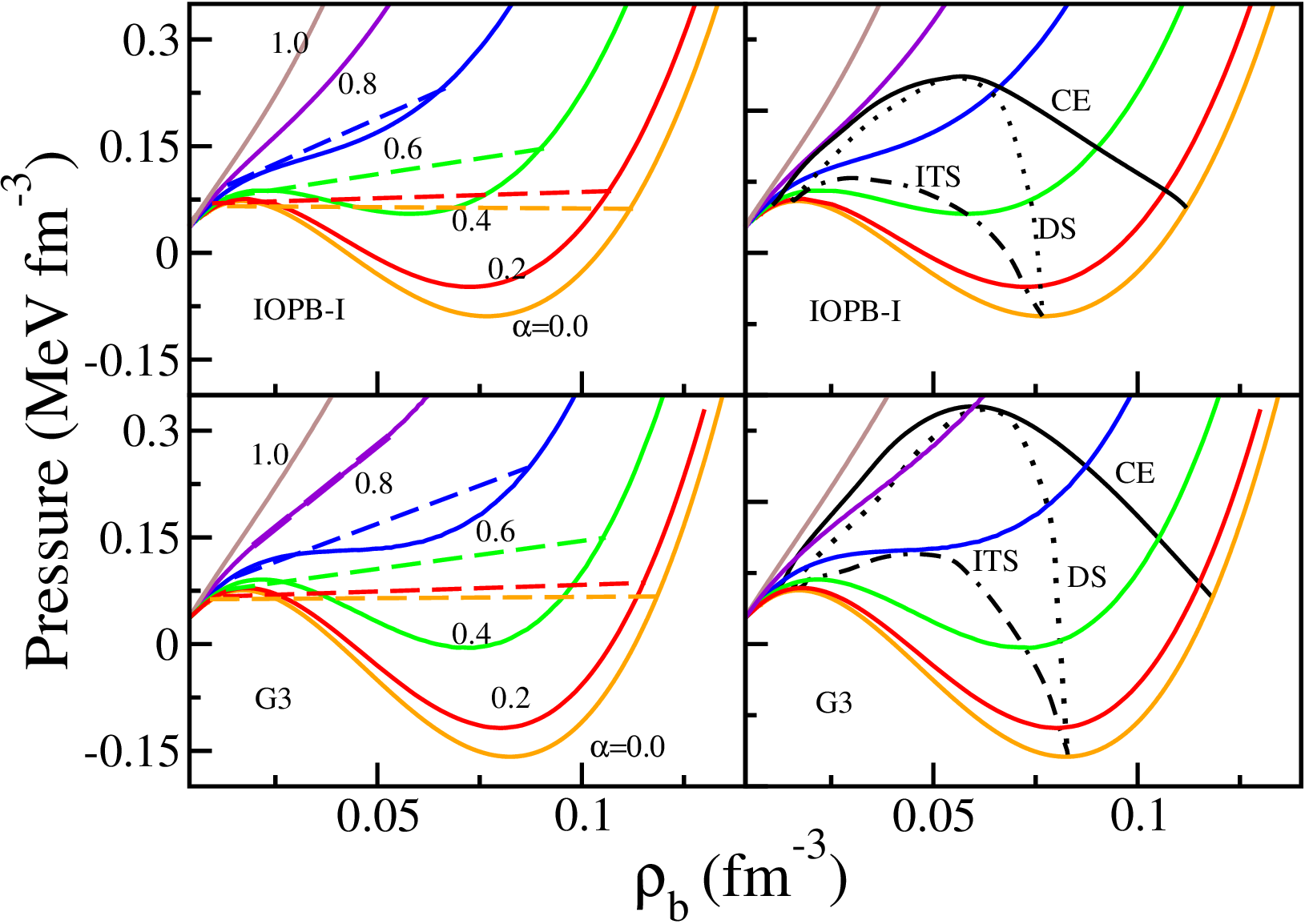}
	\caption{EoS of nuclear matter at various $\alpha$ at T= 10 MeV along with the ITS, DS and CE curves for IOPB-I and G3 sets.}
	\label{eos}
\end{figure}
The complexity of phase transition in the asymmetric nuclear matter is shown in Fig \ref{eos}. As one move from symmetric to asymmetric matter, a new behaviour distinct to the two-component system is allowed. Asymmetry is held constant during the phase transition which forces the system to change its chemical potential and consequently the pressure (shown by dashed line in left panel). Due to charge fluctuation during this phase transition, the diffusive instability appears and plays more important role than mechanical instability in describing the phase transformation. The right panel shows all  three curves i.e. ITS, DS and CE and it is visible that diffusive instability has larger area as compared to mechanical instability.
\begin{figure*}
  \centering
\subfloat[]{%
  \includegraphics[height=7cm,width=.49\linewidth]{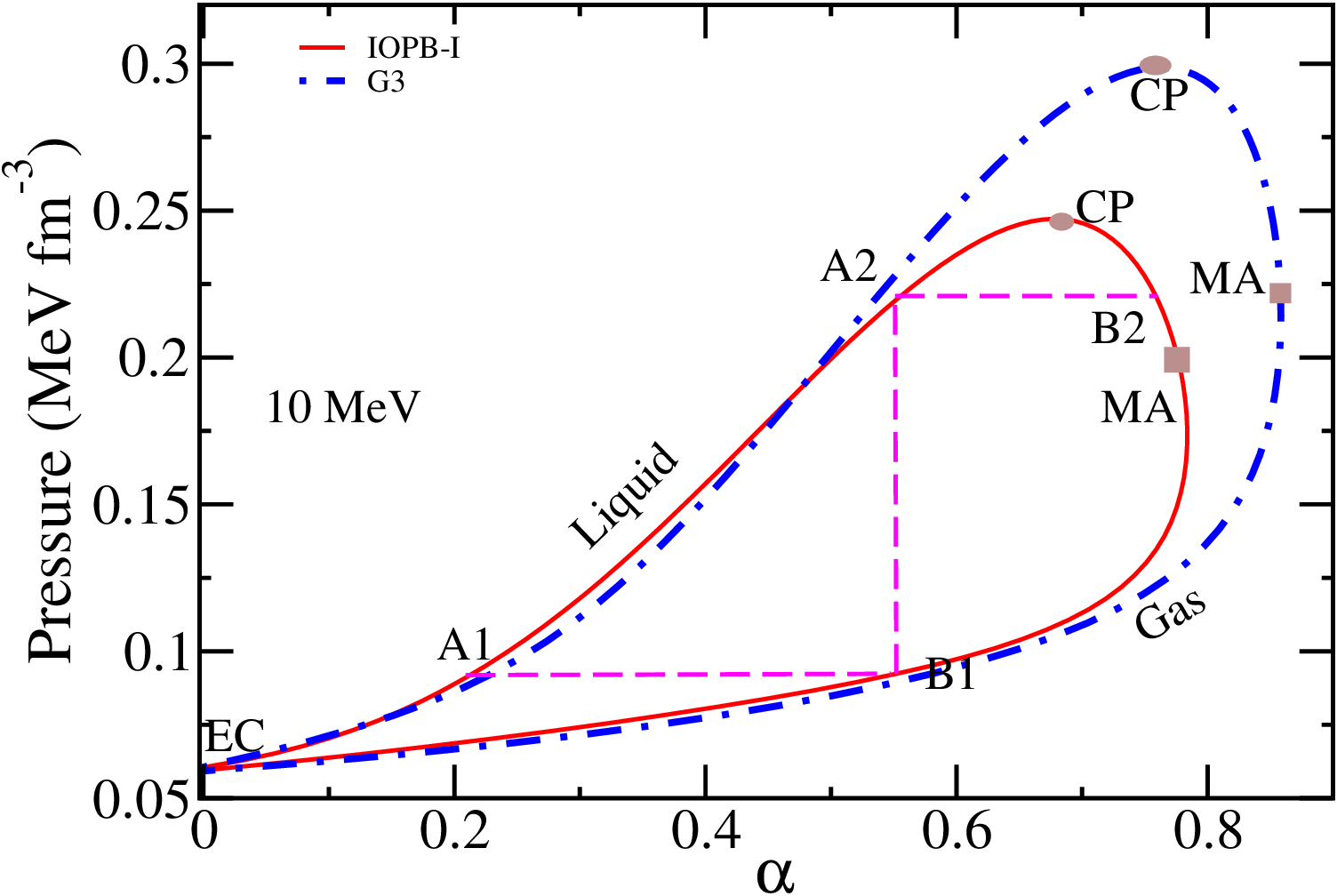}%
}\hfill
\subfloat[]{%
  \includegraphics[height=7cm,width=.49\linewidth]{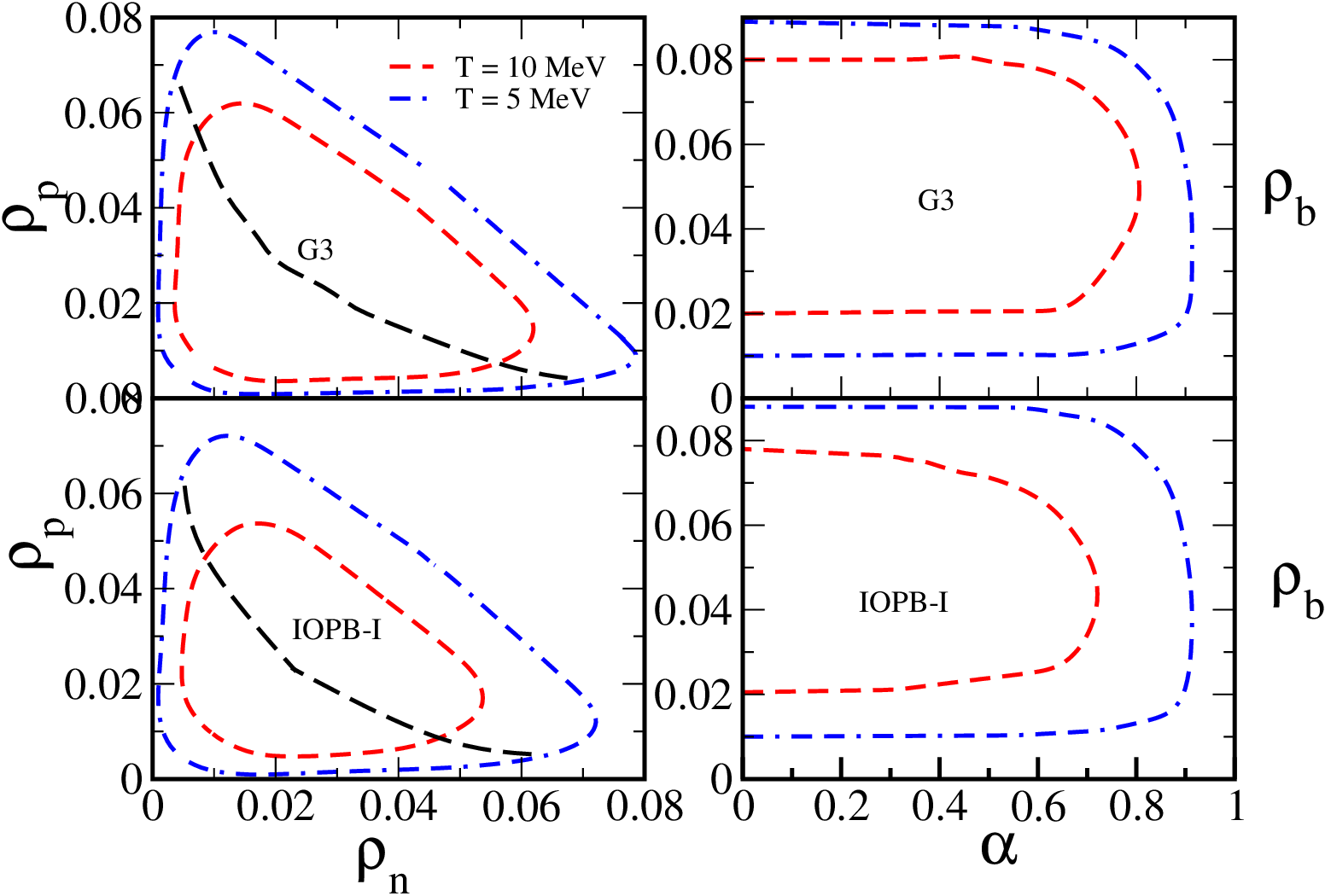}%
}
 \caption{The binodal surface on the P-$\alpha$ plane is shown on the left (a). and spinodal boundary in $\rho_n$-$\rho_p$ plane and $\alpha$-$\rho_b$ plane is shown on the right (b). The black dashed line on the spinodal shows the critical points.}
  
\label{binodalspinodal}
\end{figure*}

Binodal as per the Gibbs condition given in Eq.  \ref{gibbscondition} at T=10 MeV is plotted in Fig  \ref{binodalspinodal} by geometrical construction where a rectangle is drawn on the chemical potential isobars of neutrons and protons \cite{lattimer1978neutron}. It is characterised by the point of equal construction (EC), The point of maximal asymmetry (MA) and the critical point which determine the edge of instability area. In the phase coexistence region, the proton fraction of two-phase changes (a unique feature of two-component system) and the phase with higher asymmetry exhibits a lower density or vice-versa. At the critical temperature of symmetric matter, all the three points (EC, MA, CP) coincide and the surface becomes a point. The vertical dashed magenta line indicates that during the phase transition, $\alpha$ remain constant and both phases follow different paths i.e. liquid follow the path A1-A2 while gas phase evolves from B1 to B2 during the isothermal compression. Finally the system leave the instability at A2. This condition is called stable condensation. On the other hand, when the system is prepared with $\alpha> \alpha_c$ ($\alpha$ at CP), it operates in the gaseous phase only and this unique phenomenon is called retrograde condensation. The spinodal according to Eq. \ref{instabilitycondition} is plotted on the right side of Fig. \ref{binodalspinodal} on both $\rho_n$-$\rho_p$, $\alpha$-$\rho_b$ plane.  Fig .\ref{eos} and \ref{binodalspinodal} provide a complete description of phase transition in asymmetric nuclear matter.

In symmetric nuclear matter, the compressibility is the deciding factor for critical parameters of phase transition whereas, the phase transition in the asymmetric matter is characterised by symmetry energy. This can be verified from Eq. \ref{fsym} where the contribution of iso-spin asymmetry is reflected from the free symmetry energy $F_{sym}$ and its slope $L_{sym}$. $K_{sym}$ and $Q_{sym}$ are the higher order derivative of FNSE in the Taylor series which are still not well constrained. We have used two E-RMF sets IOPB-I and G3 to account for the various EoS properties on the phase transition in the asymmetric matter. The detailed analysis of phase transition on the symmetric matter using the IOPB-I and G3 set is discussed in \cite{vishal2020}. For ANM, the asymmetry in density is introduced by $\rho$ meson and is dictated by cross coupling $\lambda_\omega(R^2W^2)$. The G3 and IOPB-I set has $\lambda_\omega$= 0.038 and 0.024 respectively. The corresponding vales of $J$ and  $L$ at T=0 MeV are given in Table \ref{bulkproperties} whereas their finite temperature dependence is shown in Fig \ref{symenergy}. The G3 set has an additional mass asymmetry introduced by $\delta $ meson. The $\delta$ meson allows one to vary the  $L_{sym}$ without altering the symmetry energy $F_{sym}$. At a given temperature, the G3 set has larger coexistence area and large values of CP and MA as compared to the IOPB-I set sue to  the $\delta$ meson. A large coexistence area favours highly asymmetric gas in coexistence with less asymmetric dense fluid. This has a direct consequence for the core-crust transition and crust structure of neutron star. Opposite to SNM, where $\zeta_0$ plays the determining role, the value of $\lambda_{\omega}$ decide the ANM which in turn affect the $L_{sym}$. A greater $\lambda_\omega$ usually gives smaller $L_{sym}$ and vice-versa. 

\begin{figure}[h]
	\centering
		\includegraphics[width=1\linewidth]{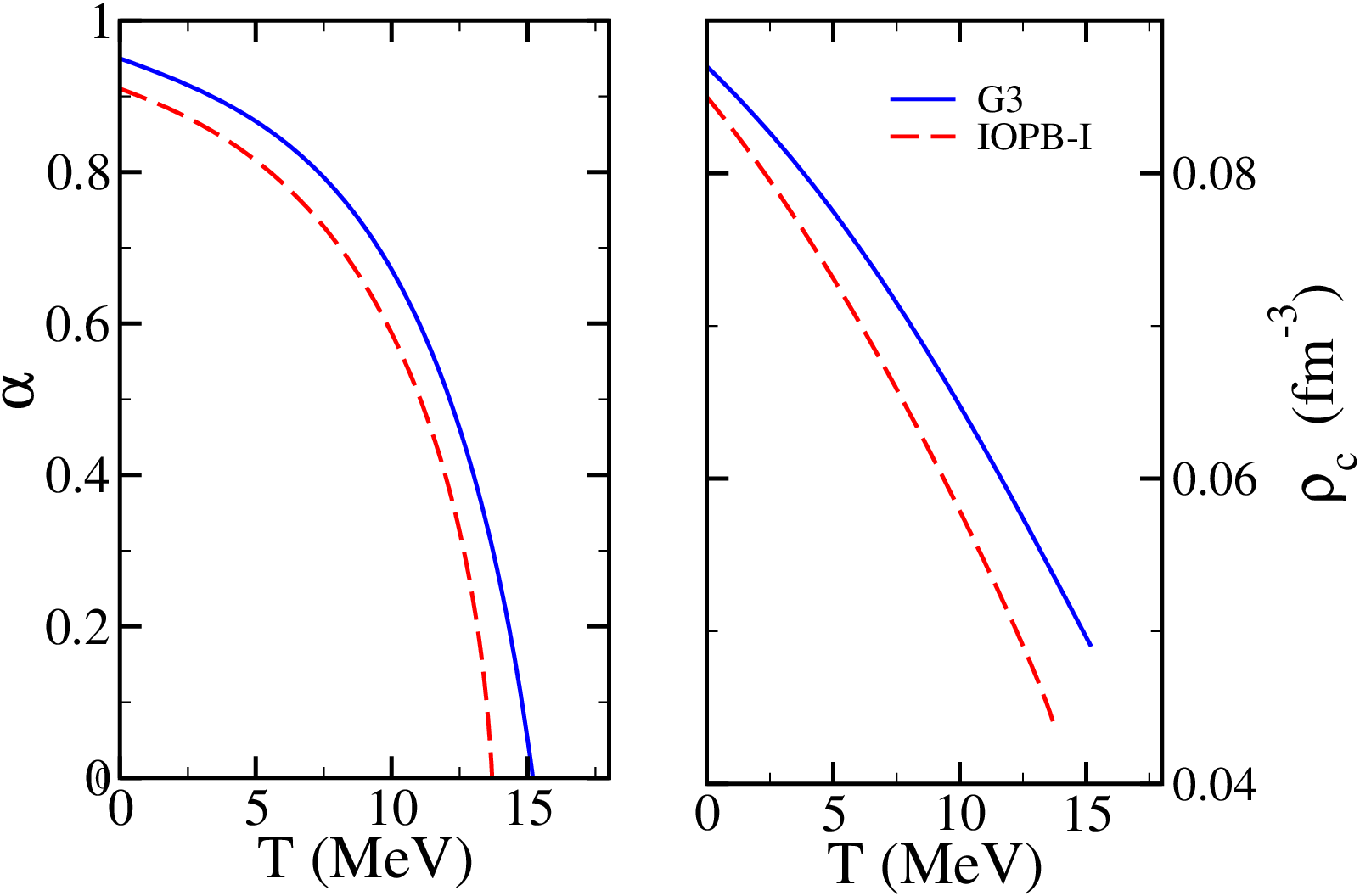}
	\caption{ $\alpha$ as a function of temperature and corresponding $\rho_c$ for the IOPB-I and G3 set.}
	\label{criticalvalue}
\end{figure}

The spinodal for the G3 set also has a larger area at any given temperature as compared to the IOPB-I set. One can observe the major variation among two sets in the coexistence densities in the $\alpha-\rho_b$ plane.  This means that the densities where different structure in non-homogeneous phase occur will be different. This property is again determined by the $L_{sym}$. The G3 set with smaller $L_{sym}$ estimates the larger $\alpha$ and $\rho_c$ at any given temperature. This is shown in Fig \ref{criticalvalue}
 where the dependence of $\alpha$ and $\rho_c$ is shown on temperature. The $\alpha-T$ plots signify the temperature at which the diffusive instability disappears (also called critical temperature). This critical temperature is not similar to symmetric matter where mechanical instability decides the phase transition but is determined according to $\frac{\partial
  \mu_p}{\partial \alpha}|_{P,T} = 0$ and  $\frac{\partial^2
  \mu_p}{\partial \alpha^2}|_{P,T} = 0$. In the E-RMF sets with constant couplings, the inflation point for proton and neutron coincides having synchronous behaviour. This might not be the case with density-dependent coupling sets \cite{Fedoseew2015, pring}. $\alpha$ decreases smoothly at low temperatures but after $T>0.5T|_{\alpha=0}$, there is a steep fall in the $\alpha$. The G3 set estimate larger $\alpha$ at a particular T due to its smaller value of $L_{sym}$ and greater value of $\lambda_\omega$. This same trend is observed in the $\rho_c$. These trends are similar to the references \cite{Alam2017, SHARMA2020121974}, where any one coupling in a parameter set were varied keeping other fixed to obtain different $L_{sym}$. The agreement of those trends while comparing two different parameter sets with almost same symmetry energy indicated that the correlation between different  properties of phase transition still holds as in case of SNM \cite{louren2017}and these can be exploited to constraint the EoS which do not take critical temperature into the account \cite{vishal2020,yang2019}.

\subsection{ \label{electron} Effect of electrons}
 In a physical system, the electrons are present so that Coulomb energy do not diverge. They are are included in EoS as a free non-interacting relativistic Fermi gas described by \cite{Avancini2006}

\begin{figure}[b]
	\centering
		\includegraphics[width=0.9\linewidth]{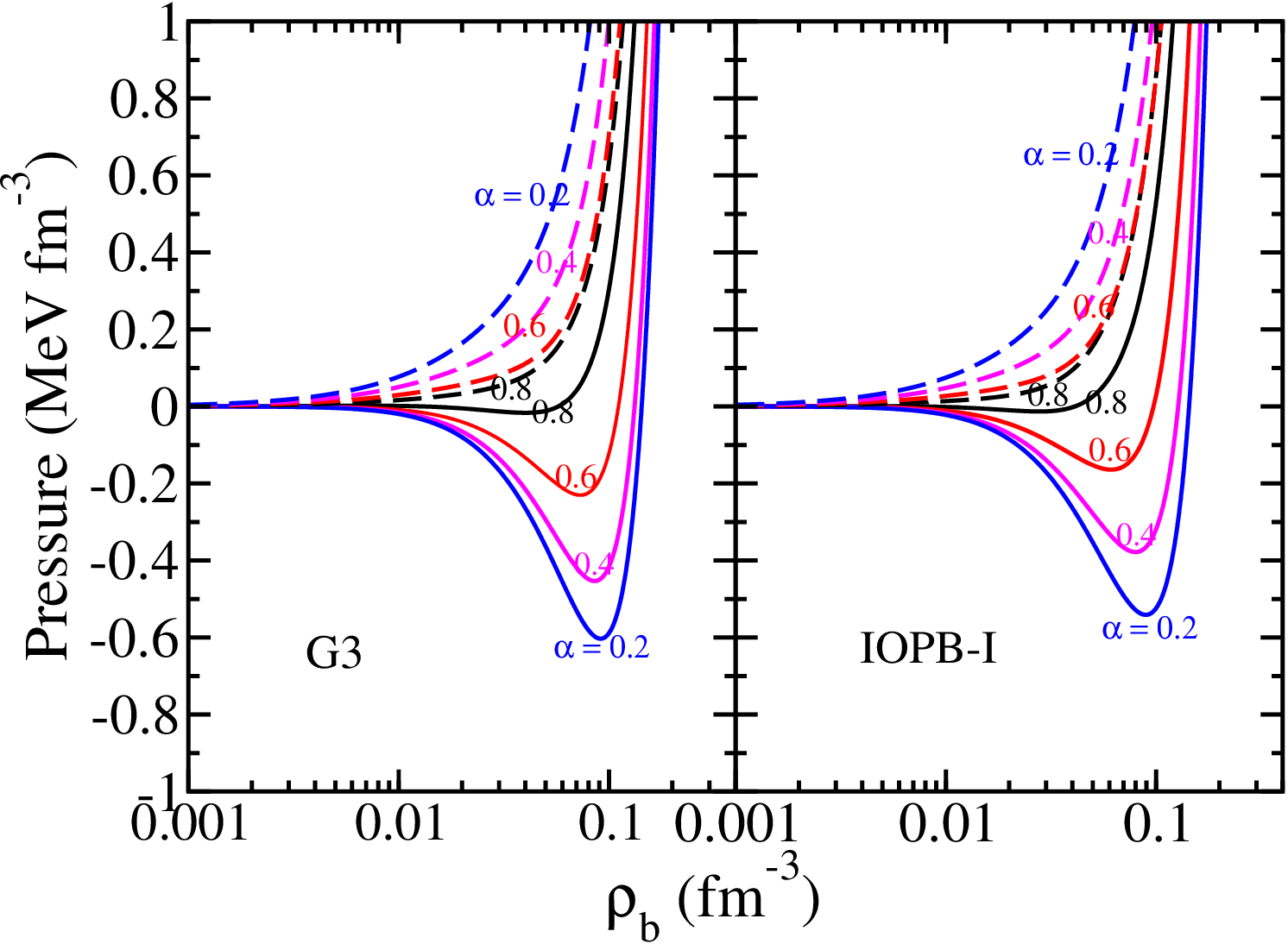}
	\caption{The EoS with and without the electrons. Solid line represents the nuclear matter without electrons and dashed line represents with electrons.}
	\label{eoselectron}
\end{figure}
\begin{equation}
L_e =  \bar{\psi_e}[i\gamma_\mu \partial^\mu - m_e]\psi_e,
\end{equation}
where $L_e$ is the Lagrangian, $m_e$ is the mass of electron. Since the electrons only compensate the  proton charge, we have $\rho_p$=$\rho_e$=$\frac{1}{\pi^2}\int k^2 dk (n_{ke}-\bar{n}_{ke})$. Where $n_{ke}$ and $\bar{n}_{ke}$ are the Fermi integral for electrons and positrons. Fig. \ref{eoselectron} shows the effect of electrons on the EoS for IOPB-I and G3 parameter sets at T=0 MeV. The effect of electrons is dominant for the matter with less asymmetry as the electron density becomes high to compensate for the larger proton density. Electrons are taken as non-interacting particles and therefore the underlying nature of a parameter set in unaltered.   Electrons have high Fermi energy and therefore, makes the system devoid of the instability. Both the IOPB-I and G3 sets have no spinodal when electrons are included for T= 5MeV. No spinodal means that stellar matter at $\beta$ equilibrium will be uniform at temperature above 5 MeV \cite{Avancini2008}. This is in consistent with the various calculations of neutron star core-crust transition. 

To further understand the implication of electrons in the EoS, we study the adiabatic index. In the processes like supernovae explosions and neutron stars, the compression and rarefactions modes of vibration are adiabatic or isentropic instead of isothermal \cite{Constantinou2015} . The adiabatic index is related to the stiffness of EoS and is given by 

\begin{equation}
    \Gamma_s=\frac{\rho_b}{P}\frac{\partial P}{\partial \rho_b} \Big|_s.
\end{equation}
\begin{figure}[h]
	\centering
		\includegraphics[width=1\linewidth]{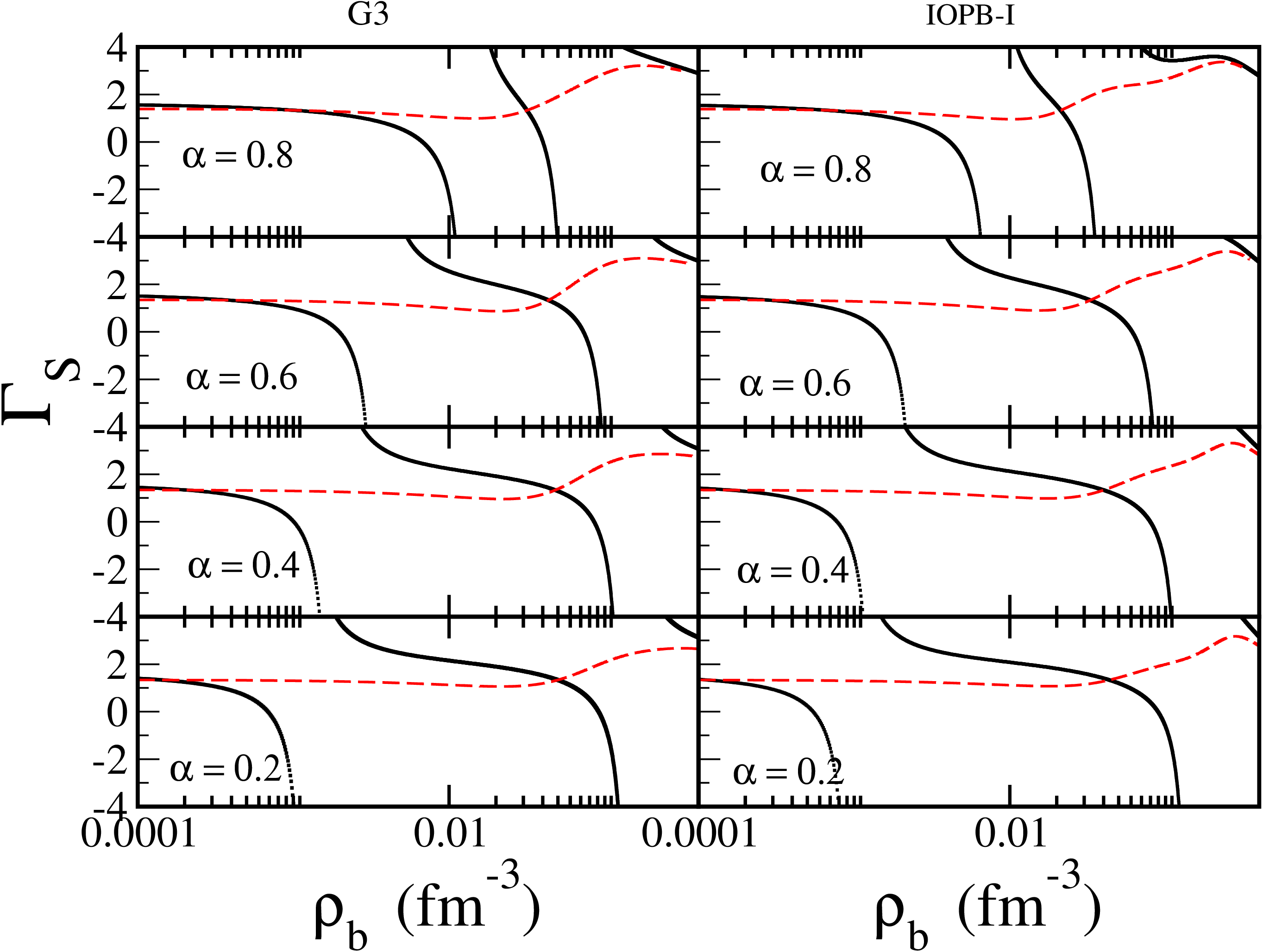}
	\caption{Adiabatic index $\Gamma_{s=0}$  for the G3 and IOPB-I set with various asymmetry. Solid black line include only nucleons while red line indicates the presence of electrons. }
	\label{adiabaticindex}
\end{figure}
$\Gamma_{s=0}$ for the two models employed here is shown in 
Fig. \ref{adiabaticindex}. The solid black curve represents the nucleon only while the red dashed curve include the contribution from electrons. $\Gamma_{s=0}$ corresponding to nucleons goes negative in some density regions showing the mechanical instability. For low and high densities it varies asymptotically. The inclusion of electrons restore the mechanical instability and value of $\Gamma_{s=0}$ increases gradually around subsaturtaion density and become asymptotically constant at low and high densities. These observations can be understood quantitatively by examining the baryon and electron pressure as shown in Fig. \ref{eoselectron}.  For $\rho \rightarrow$ 0, $\Gamma_{s=0}$ tend to the 4/3 which is due to the relativistic electrons and  is an important requirement for the stability of supernova simulation. As asymmetry rises, this value goes to 5/3 for pure neutron matter. Although the underlying properties for $\Gamma_s$ are same for both the model, but the position of instability and highest value of $\Gamma_s$ in case of matter with electrons, is essentially determined by the pressure due to baryon. The parameter set IOPB-I and G3 do not break the causality condition on speed of sound \cite{harish}.  The presence of  electron in the system also impacts the  speed of sound $C_s^2=\frac{\partial p}{ \partial E}$. Addition of electrons do not yield the nonphysical region in low density as is seen in the  nuclear matter system without leptons. At higher density, the electrons impart significant impacts on the more symmetric matter making it smoother as compared to the asymmetric matter. 

\section{\label{conclusion} Summary and Outlook}

The primary aim of this work is to study the thermal properties of hot and dilute isospin asymmetric nuclear matter within effective relativistic mean-field  (E-RMF) formalism.  Although  the thermodynamics of symmetric nuclear matter is explored, the isospin effects  are still not understood at finite temperature.  In this study, we consider the homogeneous nuclear matter and analyse it using two E-RMF sets i.e. IOPB-I and G3 at different values of  temperature and isospin asymmetry  because of their relevance in astrophysical simulations.   Both of these sets estimate the EoS at low density for pure neutron matter in agreement with microscopic chiral EFT (NN + 3N) calculations. 
Our motivation for selecting IOPB-I and G3 parameter sets  lies in the fact that these models have comparable symmetry energy at saturation but differ in the value of slope parameter $L_{sym}$ which is an important quantity for deciding the instabilities.  Furthermore, the G3 set introduces the mass asymmetry with the inclusion of $\delta$ meson which is responsible for the different effective mass of neutrons and protons which make them distinguishable. We study the  temperature dependence of free nuclear symmetry energy ($F_{sym}$) and its higher order derivative slope parameter $(L_{sym,})$, skewness parameter ($Q_{sym}$) and  curvature parameter ($K_{sym}$).  The  $F_{sym}$  increases with temperature  at a given density due to a decrease in entropy density.  The higher-order derivative of  $F_{sym}$ preserves the zero temperature behaviour with a slight change in magnitude which shows that one can use the zero-temperature value of these parameters to compare the relevant quantities at any given temperature.

To study the finite temperature effect,  we separate the thermal component from the zero temperature EoS. The thermal effects are sensitive to asymmetry at low density.  It is observed that the thermal effects in E-RMF formalism depend mainly on the density dependence of effective mass. The effective mass is calculated self consistently which depends on the  $\sigma$ and $\delta$ mesons.   A larger effective mass estimates larger thermal effects on  the state variables.  This feature is also seen in non-relativistic formalisms. 
We have calculated the isothermal thermal index ($\Gamma$) in view of its relevance  in supernovae. $\Gamma$ is very sensitive to the isospin asymmetry at a fixed temperature which is opposite to the non-relativistic case. The isothermal and isentropic incompressibility varies parabolically with temperature.  The trends of incompressibilities are in agreement with available microscopic calculations. The phase transition is studied for the asymmetric nuclear matter considering a two-component system with two conserved charges i.e. Baryon number and isospin. The G3 set due to its low $L_{sym}$ estimates the higher value of maximal asymmetry and critical pressure. The presence of $\delta$ meson has a positive effect on binding energy and therefore influences the boundary of spinodal. The critical density and asymmetry are also larger for the G3 set, which can be attributed to its lower $L_{sym}$.  The Value of $L_{sym}$ is determined mainly by cross-coupling of $\rho$ and $\omega$ meson and $\delta$ meson. One can say that a larger value of $\lambda_\omega$ estimate the larger instability in asymmetric nuclear matter.  Critical asymmetry is a quadratic function of temperature and exhibits different behaviour in the low and high-temperature range. 

Finally, we  study the effect of the electrons in EoS of nuclear matter and its instability. Electrons due to their high Fermi energy make the system devoid of instabilities. We study the adiabatic index ($\Gamma_{S=0}$) of matter with and without the inclusion of electrons. The $\Gamma_S$ with electrons become asymptotically constant at low and high densities with a small variation near the saturation density. The density of this hump predominantly depends on the baryon pressure. The electron being non-interacting particle do not alter the underlying nature of the force parameter.

The present calculations can be extended to the case of inhomogeneous matter where nuclei coexist with surrounding nucleon gas. Such studies are essential to estimate an equation of state ranging from zero density to density of inner core of neutron star.

\bibliography{reference}

\end{document}